\documentclass[pra, twocolumn, groupedaddress]{revtex4-1} 

\usepackage{epsfig}
\usepackage{graphicx}
\usepackage[normalem]{ulem}
\usepackage{xcolor}
\usepackage{amsmath, amssymb, amsthm, times, graphics, bm, bbm, dcolumn}
\usepackage[colorlinks,urlcolor=black,citecolor=blue,linkcolor=black]{hyperref}
\usepackage{multibib}
\usepackage[all]{nowidow}
\newcommand{\previous}[1]{}

\newcommand{\Li}{$^{6}$Li }
\newcommand{\ket}[1]{\ensuremath{\left|#1\right\rangle}}

\begin{document}

\title{Exploring the ferromagnetic behaviour of a repulsive Fermi gas via spin dynamics}
\author{G. Valtolina,$^{1,2,3}$ F. Scazza,$^{1,2}$ A. Amico,$^{1,2}$ A. Burchianti,$^{1,2}$\\ A. Recati,$^{4,5}$ T. Enss,$^{6}$ M. Inguscio,$^{1,2}$ M. Zaccanti$^{1,2}$ and G. Roati$^{1,2}$}

\affiliation{$^{1}$INO-CNR, Via Nello Carrara 1, 50019 Sesto Fiorentino, Italy}
\affiliation{$^{2}$LENS and Dipartimento di Fisica e Astronomia, Universit\`{a} di Firenze, Via Nello Carrara 1, 50019 Sesto Fiorentino, Italy}
\affiliation{$^{3}$Scuola Normale Superiore, Piazza dei Cavalieri 7, 56126 Pisa, Italy}
\affiliation{$^{4}$INO-CNR BEC Center and Dipartimento di Fisica, Universit\`a di Trento, 38123 Povo, Italy}
\affiliation{$^{5}$Technische Universit\"at M\"unchen, James-Franck-Stra{\ss}e 1, 85748 Garching, Germany}
\affiliation{$^{6}$Universit\"at Heidelberg, Philosophenweg 19, 69120 Heidelberg, Germany}

\begin{abstract}
\vspace*{-0.29cm}
\noindent
Ferromagnetism is a manifestation of strong repulsive interactions between itinerant fermions in condensed matter.  
Whether short-ranged repulsion alone is sufficient to stabilize ferromagnetic correlations in the absence of other effects, like peculiar band dispersions or orbital couplings, is however unclear.   
Here, we investigate ferromagnetism in the minimal framework of an ultracold Fermi gas with short-range repulsive interactions tuned via a Feshbach resonance.  
While fermion pairing characterises the ground state, our experiments provide signatures suggestive of a metastable Stoner-like ferromagnetic phase supported by strong repulsion in excited scattering states. We probe the collective spin response of a two-spin mixture engineered in a magnetic domain-wall-like configuration, and reveal a substantial increase of spin susceptibility while approaching a critical repulsion strength. Beyond this value, we observe the emergence of a time-window of domain immiscibility, indicating the metastability of the initial ferromagnetic state. Our findings establish an important connection between dynamical and equilibrium properties of strongly-correlated Fermi gases, pointing to the existence of a ferromagnetic instability.
\end{abstract}

\maketitle

\noindent The magnetic properties of a variety of quantum systems, ranging from electrons in transition metals \cite{Vollhardt2001,Brando2016} and normal $^3$He liquids \cite{VollhardtHe} to neutron and quark matter within the crust of neutron stars \cite{Silverstein1969,Tatsumi2000}, emerge from strong interactions between itinerant fermions, i.e.~not localized into a crystal lattice. Itinerant ferromagnetism can be qualitatively captured by an intuitively simple mean-field framework, as first envisioned by Stoner \cite{Stoner1933}: a free electron gas can become ferromagnetic once a short-ranged screened Coulomb repulsion between oppositely oriented electron spins overcomes the effect of Fermi pressure, which would favour a paramagnetic state with no spin ordering. A sufficiently strong repulsion promotes the parallel alignment of magnetic moments, at the price of an increased kinetic energy, making the paramagnetic state unstable towards a ferromagnetic one.
Stoner's picture has proven successful to qualitatively describe the phases of a wealth of electron systems. However, electrons in solids are subject to various effects beyond short-range repulsion \cite{Vollhardt2001,Brando2016}, which are thought to play a role in promoting or suppressing ferromagnetism. In particular, there exist materials where a Stoner-type instability of the repulsive Fermi liquid state is altered by competing mechanisms. Notable examples are the emergence of unconventional superconductivity adjoining ferromagnetism in intermetallic compounds \cite{Saxena2000} and the non-Fermi liquid behaviour of the paramagnetic state in itinerant-electron ferromagnets \cite{Pfleiderer2001}.

\begin{figure}[t!]
\begin{center}
\includegraphics[width=\columnwidth]{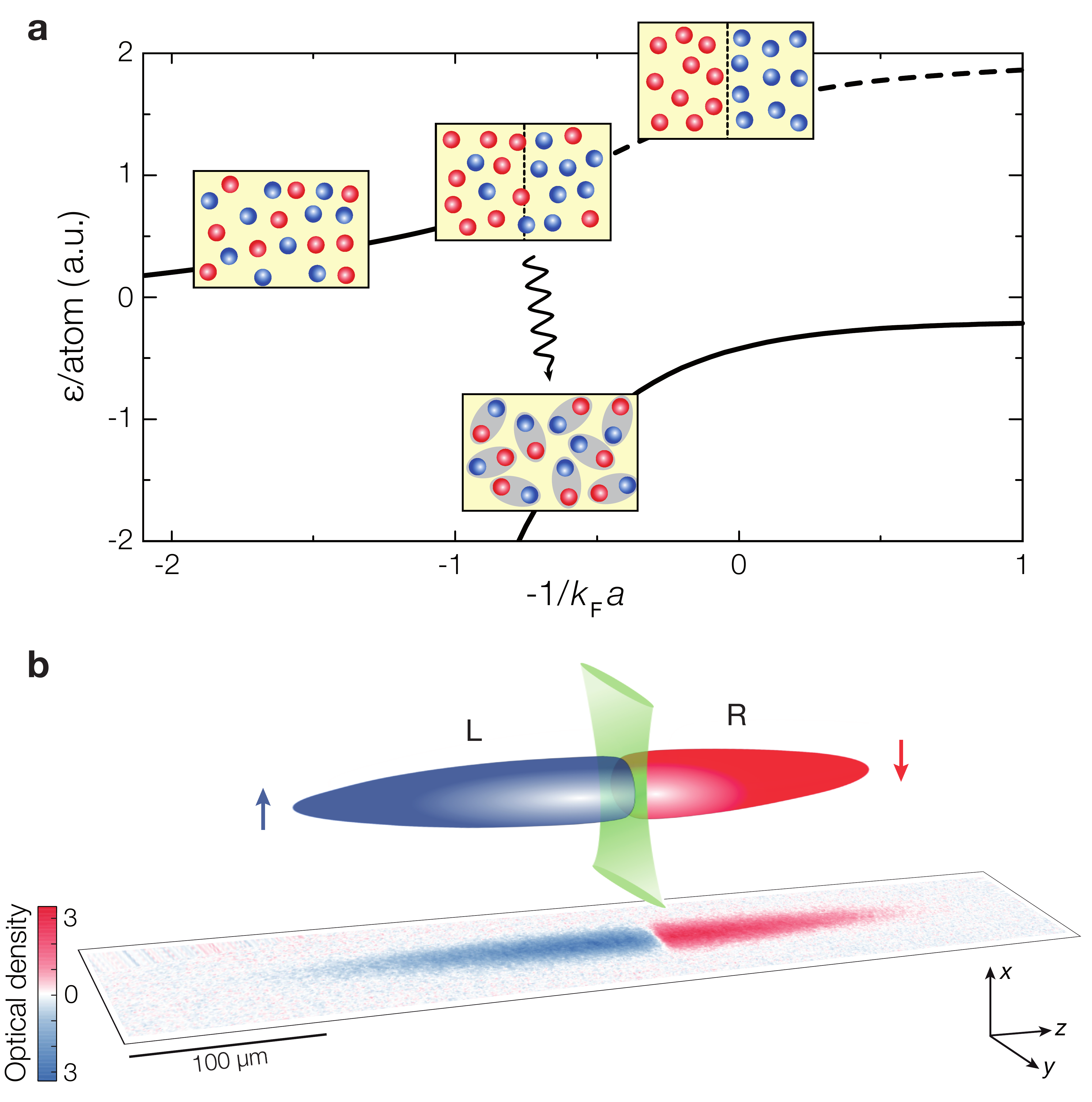}
\caption{\textbf{Engineering a ferromagnetic state with an ultracold atomic Fermi gas.} 
\textbf{a}, A Fermi mixture with resonant short-range interactions, parametrised by $1/k_F a$, features two distinct energy branches, the lower (upper) being associated to a net interspecies attraction (repulsion). Ferromagnetism may develop at strong repulsion along the upper branch, whose stability is limited due to enhanced decay processes onto the lower branch. Depending on the spin imbalance, the lower branch corresponds either to a paired phase or to an attractive Fermi liquid.
\textbf{b}, In our experiment we circumvent the pairing instability by preparing an atomic Fermi gas in a ferromagnetic domain-wall structure, probed via spin-selective \textit{in situ} imaging (see lower image). The initial state is obtained by segregating the two spin components into two initially disconnected reservoirs at equilibrium by means of a thin optical barrier with a waist of about 2\,$\mu$m (sketched in green).}
\label{Fig1}
\end{center}
\end{figure}

Quite paradoxically, even the paradigmatic scenario of an ultracold atomic Fermi gas with short-ranged repulsive interactions entails a superfluid ground state of paired fermions, rather than a ferromagnetic one. 
This stems from the fact that genuine zero-range repulsion, encoded in the $s$-wave scattering length $a$ that can be adjusted via magnetic Feshbach resonances, necessarily requires an underlying attractive potential with a weakly bound molecular state \cite{Chin2010}. Hence, the repulsive Fermi gas corresponds to a metastable excited (upper) energy branch of the many-body problem \cite{Shenoy2011, Kohstall2012, Massignan2014} (see Fig.~\ref{Fig1}a). 
It is intrinsically unstable against pair formation via inelastic decay processes \cite{Sanner2012, Kohstall2012, Lee2012} that were found to be rapid with respect to the development of magnetic domains \cite{Sanner2012, Lee2012, Pekker2011}.
Nonetheless, theoretical approaches intentionally neglecting pairing, based on Landau's Fermi liquid theory and quantum Monte Carlo (QMC) calculations \cite{Duine2005,Leblanc2009,Conduit2009,Cui2010,Pilati2010,Chang2011,Massignan2014}, confirm Stoner's ferromagnetic instability of a homogenous Fermi gas driven only by short-range repulsion. In light of the metastable nature of the repulsive branch, it is debated whether such a system may exhibit ferromagnetic correlations at all, at least temporarily, and how the magnetic and transport properties of the repulsive gas are affected by the large coupling to lower-lying states. 

Previous experiments with ultracold spin mixtures prepared in the paramagnetic phase and quenched to strong repulsion \cite{Jo2009, Sanner2012} found the Stoner instability to be precluded by the pairing one \cite{Pekker2011}, concluding that a Fermi gas with strong short-range repulsion does not undergo a ferromagnetic phase transition \cite{Sanner2012}. On the other hand, a recent spectroscopic study \cite{Scazza2016} of highly-imbalanced repulsive mixtures in the polaronic regime \cite{Kohstall2012, Massignan2014} revealed the Fermi liquid phase to become energetically and thermodynamically unstable beyond a critical repulsion strength, in quantitative agreement with theoretical predictions for polarized repulsive gases \cite{Cui2010, Pilati2010, Schmidt2011, Massignan2014}, while still featuring an unexpectedly long lifetime before decaying onto lower-lying states. This supports a scenario where the repulsive Fermi gas may exhibit a transient time during which repulsion and ferromagnetic correlations dominate the evolution of the many-body system also in the balanced case, for which a ferromagnetic instability is maximally favoured \cite{Pilati2010}.

In this work we investigate the spin response and the stability of a balanced spin $\uparrow$ -- $\downarrow$ mixture prepared in an artificial magnetic domain-wall structure (see Fig.~\ref{Fig1}b). Such a fully-ferromagnetic initial configuration features a vanishingly small $\uparrow$ -- $\downarrow$ density overlap in contrast to a paramagnetic state \cite{Jo2009, Sanner2012}. This greatly suppresses the effect of pairing processes, ensuring that the overall relaxation rate remains much slower than that given by the Fermi energy, thereby maintaining the system on the upper branch for a comparatively long time. 
We probe the ferromagnetic properties of the repulsive Fermi gas through two distinct but interconnected measurements of spin dynamics at varying interaction. On the one hand, by probing the spin-dipole mode \cite{Recati2010}, i.e.\,\,the out-of-phase relative oscillation of the approaching spin $\uparrow$ -- $\downarrow$ domains and observing its softening, we 
demonstrate an increase of the spin susceptibility as a function of repulsion strength, up to a sharp discontinuity in the spin response. On the other hand, by studying spin diffusion \cite{Sommer2011,Enss2012, Bardon2014} at short and long evolution times we reveal the metastability of the initial ferromagnetic state and gain new insights on its relaxation mechanisms \cite{Sanner2012, Pekker2011, Massignan2014}. In particular, our work shows that the short-time collective dynamics of an artificially-created ferromagnetic state are governed by strong repulsion, before inelastic decay to the attractive branch eventually leads to demagnetization. 

We initially prepare a weakly interacting mixture of ultracold $^6$Li atoms \cite{Burchianti2014}, equally populating the two lowest Zeeman states, hereafter denoted as $\left | \uparrow \right\rangle$ and $\left | \downarrow \right\rangle$. The atoms are held in a cylindrical optical dipole trap with axial and radial frequencies $\nu_{z}\simeq 21$ Hz and $\nu_{\bot}\simeq 265$ Hz, respectively. 
By adjusting the evaporation procedure we can tune the degree of degeneracy from $T/T_F < 0.1$ up to $\sim 1$. Here $T$ is the gas temperature, while $T_F$ is the Fermi temperature of a single-component Fermi gas of $N$ atoms in a harmonic trap, given by $k_B T_F=E_F= h (6 N \nu_{z} \nu_{\bot}^2)^{1/3}$, with $h$ and $k_B$ denoting the Planck's and Boltzmann's constants.
At a magnetic field of about 1\,G, where the magnetic moments of $\left | \uparrow \right\rangle$ and $\left | \downarrow \right\rangle$ states are opposite, the application of a magnetic field gradient allows us to spatially separate the two spin components along the weak axis of the trap. Once the overlap between the two clouds is perfectly zero, we superimpose a 2 $\mu$m thin optical repulsive barrier as high as $V_0\sim 10 E_F$ onto the center of the harmonic potential, in order to split the trap into two independent reservoirs \cite{Valtolina2015}, as sketched in Fig.~\ref{Fig1}b. We then adiabatically turn off the magnetic field gradient and end up with all $\uparrow$ ($\downarrow$) fermions in the left (right) reservoir (see Methods and Supplementary Information for details).
This creates two macroscopic spin domains at rest, separated by a distance only a few times wider than the mean interparticle spacing of the gas. Such configuration resembles the density distribution expected for a spin-mixture at full magnetization in an elongated harmonic trap, with a central domain wall of thickness around the interparticle spacing, additionally surrounded by an unpolarized low-density shell on the cloud surface.
From here, we let the two spin components start interacting by removing the optical barrier. Before switching off the barrier, the interaction strength is adjusted by setting the magnetic field close to the center of a broad $\uparrow$ -- $\downarrow$ Feshbach resonance located around 832 Gauss \cite{Zurn2013}. 

\begin{figure*}
\begin{center}
\includegraphics[width=12cm]{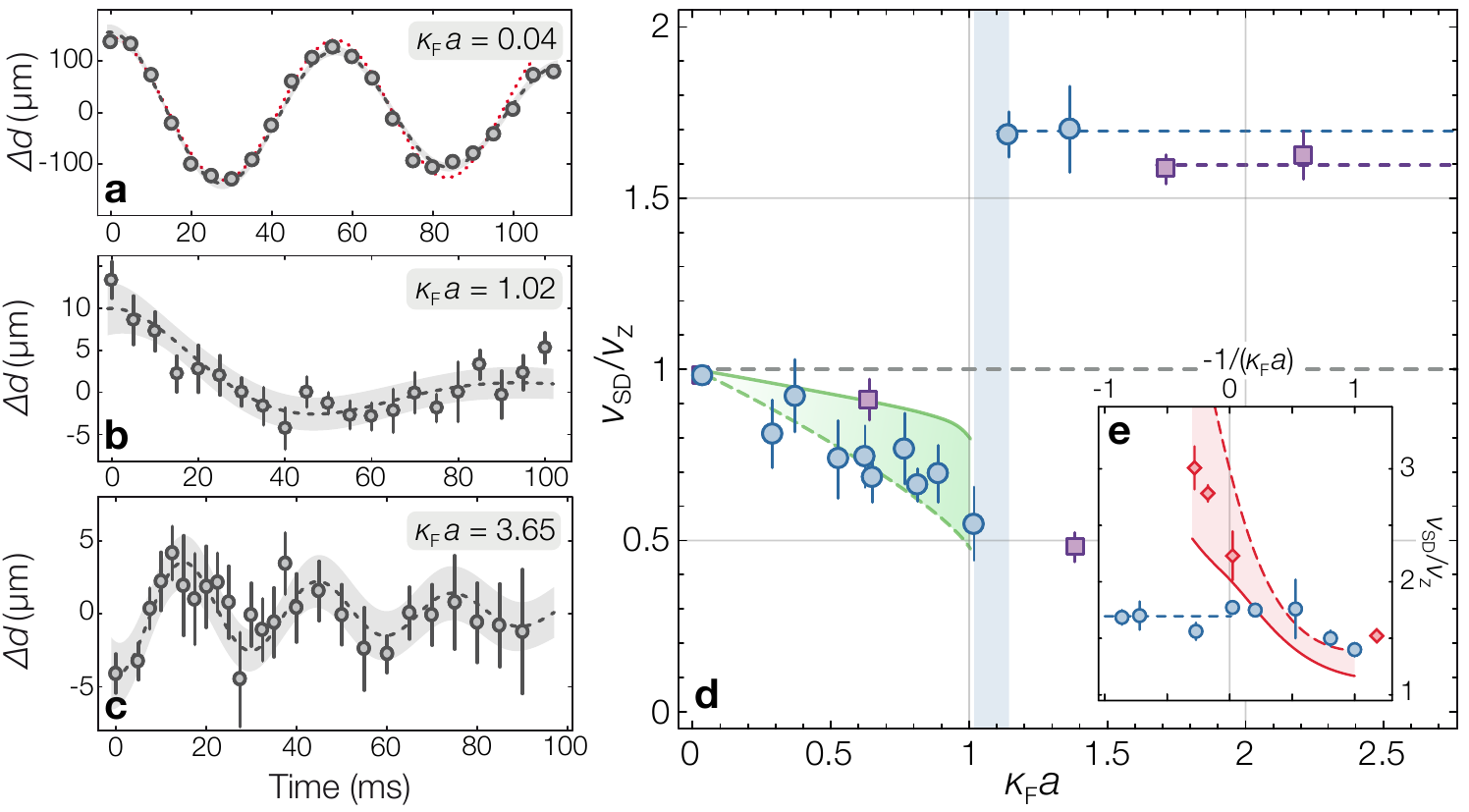}
\caption{\textbf{Spin response of a repulsive Fermi gas.} 
\textbf{a-c}, After subtracting a slow exponential drift from the centre-of-mass distance $d(t)$ between the two spin clouds, the residual out-of-phase dynamics $\Delta d(t)$ after sudden barrier removal is fitted to a damped sinusoidal function (dashed lines), from which $\nu_{SD}$ is extracted for several interaction strengths. The bare trap oscillation is shown for comparison (dotted red line in \textbf{a}). Shaded areas denote the standard confidence bands of the fits. Data points result from at least 5 independent measurements with error bars given by the standard error of the mean (s.e.m.) combined with the uncertainty on the subtracted exponential drift. \textbf{d}, The normalized spin-dipole frequency $\nu_{SD}/\nu_z$ is plotted versus $\kappa_F a$ for $T/T_F=0.12(2)$ (blue circles) and $T/T_F=0.25(4)$ (purple squares), with error bars being 95$\%$ confidence intervals of sinusoidal fits. A decrease of $\nu_{SD}$ followed by a clear discontinuity is visible, 
 suggesting the occurrence of a ferromagnetic instability.
Here, the shaded blue interval denotes the range of interaction strengths where the spin-dipole mode does not exhibit a distinct single-frequency oscillatory behaviour. The dashed blue (violet) lines are the average $\nu_{SD}/\nu_z$ measured beyond the 
discontinuity point at $T/T_F \simeq 0.12$ ($T/T_F\simeq 0.25$) up to unitarity. The solid (dashed) green lines are the $T=0$ predictions from a sum-rule approach assuming 25$\%$ (100$\%$) $\uparrow$ -- $\downarrow$ spatial overlap (see Methods). \textbf{e}, $\nu_{SD}/\nu_z$ measured at $T/T_F \simeq 0.12$ across the Feshbach resonance is shown for the gas initially prepared in the upper branch (blue circles) or in the lower branch (red diamonds). The solid (dashed) red line represents the spin-dipole frequency prediction using the attractive gas spin susceptibility from Ref.~\citenum{Tajima2016} and setting a 20$\%$ (100$\%$) $\uparrow$ -- $\downarrow$ spatial overlap (see Supplementary Information). The dashed blue line coincides with the one shown in panel \textbf{d}.}
\label{Fig2}
\end{center}
\end{figure*}

\bigskip\noindent\textbf{Spin response of a repulsive Fermi gas}\\In a first experiment the spin dynamics is triggered by abruptly switching off the barrier from its initial value on a $\mu$s timescale. Owing to the small initial separation of about 5\,$\mu$m, the two spin clouds approach each other with small relative momentum $\hbar k \ll \hbar k_F = \sqrt{2 m_\text{Li} E_F}$. We follow the clouds dynamics by resonant \textit{in-situ} absorption imaging, monitoring the evolution of the two spin domains (see Methods and Supplementary Information). On top of an overall slow drift, the relative distance between their centres of mass $d(t)=z_{\uparrow}(t)-z_{\downarrow}(t)$ presents a small-amplitude out-of-phase oscillation, signalling the excitation of the spin-dipole mode (see Fig.~\ref{Fig2}a-c).
The measurement of the spin-dipole frequency\cite{Recati2010,Bienaime2016}, next to the one of spin fluctuations \cite{Sanner2011,Sanner2012}, is a powerful probe for disclosing the magnetic properties of harmonically trapped gases, connecting the local behaviour of the spin susceptibility $\chi$ to a directly accessible quantity.
Following a sum-rule approach \cite{Recati2010}, the spin-dipole frequency $\nu_{SD}$ is found to be:
\begin{equation}
	\nu_{SD}^2=\frac{N_{\uparrow}+N_{\downarrow}}{4 \pi^2 m_\text{Li} \int d\textbf{r} z^2 \chi(n(\vec{r}))},
\label{nuSDweightedMain}	
\end{equation}
where $\chi$ depends on the local density $n(\vec{r})$ in the trap (see Supplementary Information for details). In particular, for a non interacting gas $\nu_{SD}=\nu_z$.
An increase of $\chi$ in the repulsive Fermi liquid phase can therefore be distinctly identified by $\nu_{SD} < \nu_z$, i.e. by a softening of the spin-dipole mode. Although $\chi$ is locally divergent if a ferromagnetic instability is reached at the trap center, $\nu_{SD}$ can only reach a non-zero minimum \cite{Recati2010} owing to the inhomogeneous density profile of the trapped gas, whose outer low-density paramagnetic region contributes with a small spin susceptibility.
The measurement of this collective oscillation is extremely challenging when starting from a paramagnetic configuration due to strong damping \cite{Duine2010} and inelastic processes \cite{Sanner2012,Kohstall2012,Scazza2016}. 
Here, instead, where the two spin domains just partially overlap over the timescale of the measurement, we are able to trace a few oscillation periods of the spin-dipole mode from which we extract $\nu_{SD}$ through a fit to a damped sinusoidal 
function (see Supplementary Information). Performing such a measurement at several magnetic field values, we obtain the trend of the spin-dipole frequency as a function of the repulsive interaction strength, displayed in Fig.~\ref{Fig2}d for two distinct temperatures. 
The interaction strength is described by the dimensionless parameter $\kappa_F a$, where $\kappa_F$ (and correspondingly $\epsilon_F$) is the average Fermi wave number (energy) weighted over the initial density distribution close to the interface between the two domains (see Methods). 

Let us discuss here the results for the colder samples at $T/T_F = 0.12(2)$ (blue circles in Fig.~\ref{Fig2}d). By starting from the weakly interacting regime, where $\nu_{SD} \simeq \nu_{z}$ (see Fig.~\ref{Fig2}a), an increase of the interspecies repulsion leads to a progressive reduction of the spin-dipole frequency, down to values as low as $\nu_{SD} \simeq 0.6\,\nu_{z}$ at about $\kappa_F a \simeq 1$ (Fig.~\ref{Fig2}b). We find the decrease of $\nu_{SD}$ to be accompanied by a strong increase of the damping of the oscillations \cite{Duine2010}. 
By further increasing the interspecies repulsion, an abrupt change occurs in the spin dynamics: for $\kappa_F a \gtrsim 1.1$, the spin-dipole frequency sharply jumps above the bare trap frequency, $\nu_{SD} \simeq 1.70(4)\,\nu_z$ (see Fig.~\ref{Fig2}c), while the damping of the oscillations is strongly reduced. Once this narrow interaction region is crossed, a further increase of $\kappa_F a$ does not produce any significant change, neither in the damping rate nor in $\nu_{SD}$. Higher temperature data exhibit a qualitatively similar trend, with the abrupt change in $\nu_{SD}$ occurring at a higher $\kappa_F a$ value.

The observed trend of $\nu_{SD}$ up to $\kappa_F a \simeq 1$ agrees with the mode-softening predicted by Eq.~\eqref{nuSDweightedMain}, reflecting a substantial increase of spin susceptibility. This is further supported by the good agreement of our low-temperature data with the spin-dipole frequency trend calculated in linear response \cite{Recati2010} (see green lines in Fig.~\ref{Fig2}d), inserting into Eq.~\eqref{nuSDweightedMain} the susceptibility $\chi(\kappa_F a)$ from QMC calculations for a $T=0$ homogeneous repulsive Fermi gas \cite{Pilati2010} (see Methods). Furthermore, the value of $\nu_{SD}$ for $\kappa_F a > $ 1.1, compatible with previous measurements at unitarity \cite{Sommer2011}, compares well with the one expected for two immiscible clouds of unpaired fermions bouncing off each other \cite{Taylor2011}, as if one spin component acted like an impenetrable potential barrier for the other. The increase of spin susceptibility detected up to a critical repulsion strength together with the discontinuity of the spin response frequency hint at a ferromagnetic instability located at $\kappa_F a \approx 1$ for $T/T_F \simeq 0.12$.

The measured behaviour of $\nu_{SD}$ could not be explained if pairing processes dominated the system dynamics. On the lower branch, owing to effective attraction, the spin susceptibility decreases monotonically while crossing the Feshbach resonance from $a<0$ to $a>0$, vanishing completely for a Bose-Einstein condensate (BEC) of pairs. This trend equally occurs in the attractive Fermi liquid state \cite{Nascimbene2011} as well as in the superfluid or in the elusive pseudo-gap state \cite{Sanner2011,Tajima2016}. 
Therefore, Eq.~\eqref{nuSDweightedMain} implies a corresponding monotonic increase of $\nu_{SD}$ moving towards the BEC limit. We experimentally demonstrate this behaviour by intentionally initializing the system in the lower branch (see Supplementary Information), observing an increase of $\nu_{SD}$ even above 2$\nu_{z}$ for $\kappa_F a > 0$ (see Fig.~\ref{Fig2}e). The spin response measured for the attractive gas is consistent with predictions from Eq.~\eqref{nuSDweightedMain}, obtained by inserting the spin susceptibility $\chi$ recently calculated for a BEC-BCS crossover Fermi gas above the critical temperature for superfluidity \cite{Tajima2016}, in agreement with previous measurements \cite{Sanner2011}. In particular, it qualitatively differs from the spin response at $\kappa_F a > 0$ shown in Fig.~\ref{Fig2}d, leading us to exclude that the latter arises from paired fermions in the interface region. On the other hand, to confirm that during the spin dynamics shown in Fig.~\ref{Fig2}a-c the upper branch is indeed predominantly occupied, we estimate the molecular fraction under the same experimental conditions by an independent measurement (see Supplementary Fig.~S6). We detect a molecular fraction not exceeding 15\% after the first 50\,ms of evolution for any $\kappa_F a$ value herein investigated, and below 5\% for $\kappa_F a \gg 1$. Additionally, our findings seem incompatible with a purely collisional picture \cite{Goulko2011} (see Supplementary Information), also considering the markedly distinct behaviour of $\nu_{SD}$ on the attractive branch shown in Fig.~\ref{Fig2}e. 

\smallskip\noindent\textbf{Observation of spin domain immiscibility}\\If the initially prepared fully ferromagnetic state was indefinitely stable above a critical $\kappa_F a$, the two spin domains would remain immiscible, i.e.\,\,spin diffusion would be impeded \cite{Duine2010}. To investigate this aspect, we study spin diffusion in a second set of measurements, exploring various interaction and temperature regimes. 
We initialize the dynamics by adiabatically lowering the barrier height through a 30\,ms linear ramp from $V_0 \sim10\,E_F$ down to 2\,$E_F$, letting the two clouds slowly approach each other. Their relative distance is reduced from about 5\,$\mu$m down to 1\,$\mu$m, yet each spin domain remains confined within its own reservoir. At this point, we remove the barrier in 5\,ms and we monitor the subsequent evolution of the relative population of the $i=\uparrow,\downarrow$ component in the left and right reservoirs, $M_{i}=(N_{i, L}-N_{i, R})/(N_{i,L}+N_{i,R})$, from which we obtain the magnetization $\Delta M=(M_{\uparrow}-M_{\downarrow})/2$. Since the distance between the two nearby cloud edges is approximately equal to the local interparticle spacing at the interface between the two spin domains, this procedure does not excite any detectable spin-dipole oscillation.  

\begin{figure*}
\begin{center}
\includegraphics[width=0.7\textwidth]{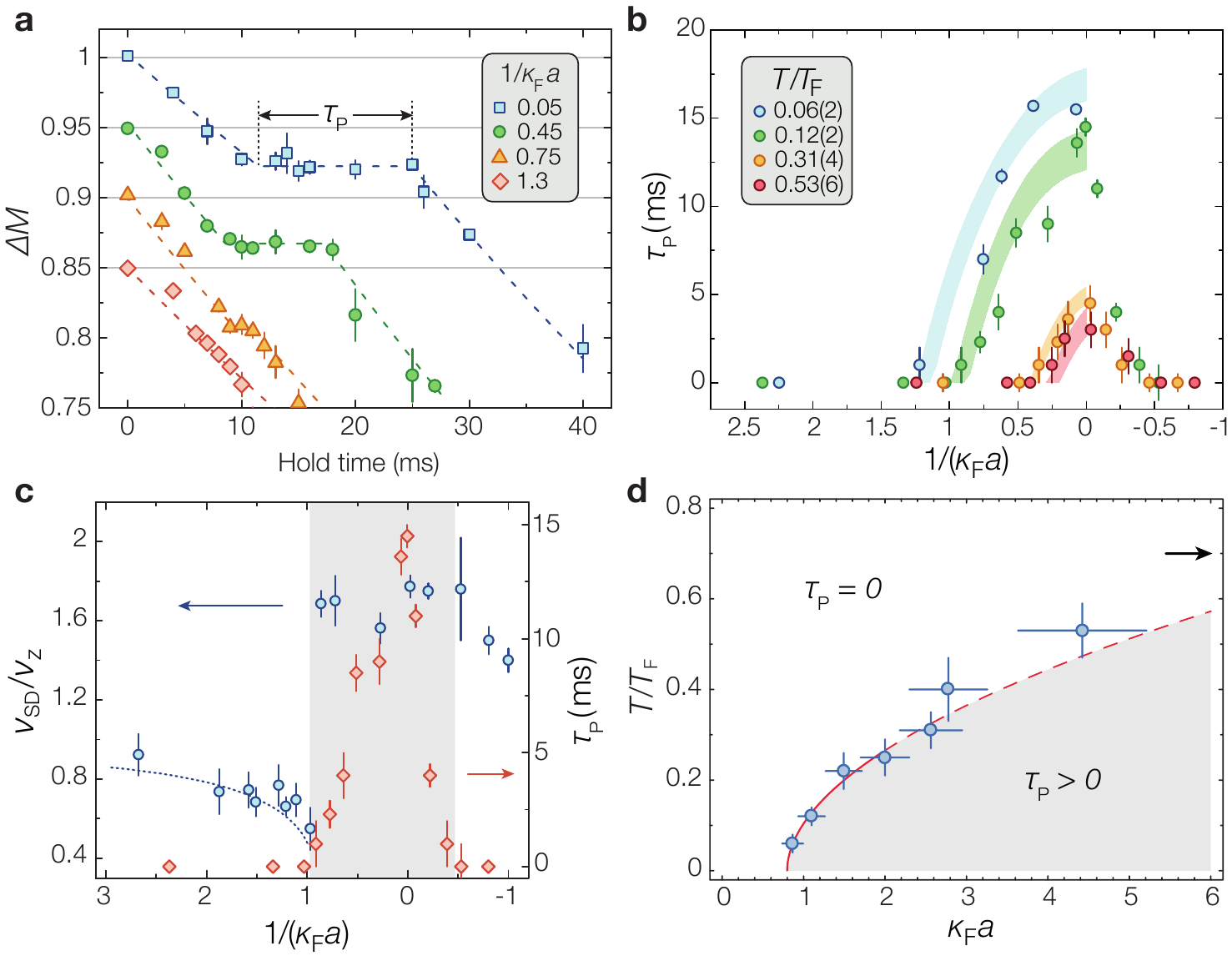}
\caption{\textbf{Metastability of a fully magnetized ultracold Fermi gas.} \textbf{a}, Evolution of $\Delta M (t)$ for different interactions at $T/T_F=0.12(2)$. For $\kappa_F a \geq 1.1$ a time window $\tau_p$ of vanishing spin diffusion is detected. Error bars are the s.e.m.\! of 4-5 independent measurements. Data sets are artificially shifted by 0.05 along the $y$-axis from one another for clarity. \textbf{b}, $\tau_p$ from piecewise fits to the data (dashed lines in \textbf{a}) is plotted at varying $1/\kappa_F a$ for various temperatures, with error bars being the fit standard uncertainty. Shaded curves are derived from the proposed model for domain-wall melting (see Methods), adjusting $E_{+c}$ within a 20\% variation. \textbf{c}, Connection between $\tau_p$ (red diamonds) and $\nu_{SD}$ (blue circles) at $T/T_F=0.12(2)$. The grey region depicts the regime of $\tau_p>0$, which coincides within the experimental accuracy to the region where $\nu_{SD} \simeq 1.7$. The dotted blue line denotes the prediction on $\nu_{SD}$ for the repulsive Fermi liquid (see Fig~2d). \textbf{d}, Metastability ($\tau_p>0$) region of the ferromagnetic state in the temperature-interaction plane. $y$-error bars denote the experimental uncertainty on $T/T_F$, while $x$-error bars account for the uncertainty on estimating the critical $\kappa_F a$ at which a $\tau_p > 0$ is observed. 
The solid line is a power-law fit (see text) to $T/T_F < 0.3$ points, extended over all values of $\kappa_F a$ as a dashed line. The black arrow marks the temperature for which $\tau_p=0$ at any $\kappa_F a$.}
\label{Fig3}
\end{center}
\end{figure*}

Figure~\ref{Fig3}a shows the short-time evolution $\Delta M(t)$ for different interaction strengths: above a critical value of repulsion, after an initial slight decrease of $\Delta M$ from 1 to about 0.92, we indeed observe a time window during which spin diffusion is completely arrested. We attribute the initial drop in magnetization to spin transport within the low-density outer shells of the two spin clouds.   
The duration of the magnetization ``plateau'' is however finite, since the stability of our ferromagnetic state is limited by the intrinsic tendency of the system to relax from the excited upper branch onto lower-lying energy states \cite{Pekker2011,Massignan2014} (see Fig.~\ref{Fig1}a). A thorough characterization of this interesting feature is summarized in Fig.~\ref{Fig3}b, where we plot the measured plateau duration $\tau_p$ of constant $\Delta M$ as a function of $1/(\kappa_F a)$, for various temperatures. 
The value of $\tau_p$ is determined through a piecewise linear fit to the data, yielding zero when no noticeable halt of spin dynamics is detected. 
A non-zero $\tau_p$ is detected only above a critical $\kappa_F a$ value and below a certain $T/T_F$. Notably, finite plateaus appear above an interaction strength nearly coincident with the one at which the spin-dipole mode frequency in Fig.~\ref{Fig2}d reaches its minimum and exhibits an abrupt change (see Fig.~\ref{Fig3}c). Furthermore, by increasing $\kappa_F a$ (increasing $T/T_F$) $\tau_p$ increases (decreases), reaching its maximum at the unitary point $1/(\kappa_F a)=0$. On the other hand, no dynamical arrest is observed for $T/T_F\ge 0.7$. In addition, at low temperatures the trends for $\tau_p$ and $\nu_{SD}$ (see Fig.~\ref{Fig3}c and Fig.~\ref{Fig2}e) suggest that we access the upper branch even within a narrow $1/(\kappa_F a)<0$ region beyond unitarity, in accord with recent predictions \cite{Pekker2011}. 

We find the behaviour of the plateau duration for  $1/(\kappa_F a) > 0$ to be captured (see curves in Fig.~\ref{Fig3}b) by a phenomenological model based only on the knowledge of the lifetime and energy spectrum of the upper and lower branches of the many-body system \cite{Schmidt2011}, calculated in the extremely polarized limit of one single $\uparrow$ ($\downarrow$) impurity embedded in a $\downarrow$ ($\uparrow$) Fermi gas \cite{Schmidt2011,Massignan2014}. Based on such a description, the ferromagnetic state is destroyed by inelastic processes occurring at the interface between the two macroscopic spin domains: fermions of one kind, overcoming the surface tension associated with a domain wall, can deposit an overall excess energy through decay from the upper to the lower branch. Only after some time, once a sufficient energy has been released into the system, the domain wall is melted and spin diffusion is established.

Upon identifying for each temperature the lowest $\kappa_F a$ value at which a non-zero $\tau_p$ of steady magnetization is observed, we delimit a region in the interaction-temperature plane where the ferromagnetic domains remain temporarily immiscible, as displayed Fig.~\ref{Fig3}d. 
The critical interaction strength for the onset of a non-zero $\tau_p$ displays a non-linear dependence upon temperature, and by fitting the $T/T_F < 0.3$ data points with $T/T_F \propto ((\kappa_F a)(T)-(\kappa_{F} a)(0))^{\alpha}$, we obtain $\alpha=0.52(5)$ and $(\kappa_{F} a)(0)=0.80(9)$. The fitted exponent matches within its uncertainty the value $\alpha=1/2$ expected from the low-temperature behaviour of a Fermi liquid exhibiting a magnetic instability (see Supplementary Information). The extracted zero-temperature value $(\kappa_{F} a)(0)$ is interestingly found in good agreement with the critical one obtained from repulsive QMC calculations \cite{Pilati2010, Chang2011}, and is significantly lower than Stoner's mean-field criterion for an unpolarized gas \cite{Stoner1933,Massignan2014, Duine2005, Leblanc2009, Conduit2009, Pilati2010} $\kappa_{F} a = \pi/2$. Notwithstanding the metastable nature of the ferromagnetic state and the dynamical character of our study, our findings agree with theoretical expectations for a repulsive Fermi gas at equilibrium undergoing a ferromagnetic instability in the absence of pairing. Moreover, the close correspondence between the trends of $\nu_{SD}$ and $\tau_p$ (see Fig.~\ref{Fig3}c) further suggests that the critical interaction strength for $\tau_p >0$ corresponds to the one required for the fully ferromagnetic state to be favoured. 

\begin{figure}[t]
\begin{center}
\includegraphics[width=\columnwidth]{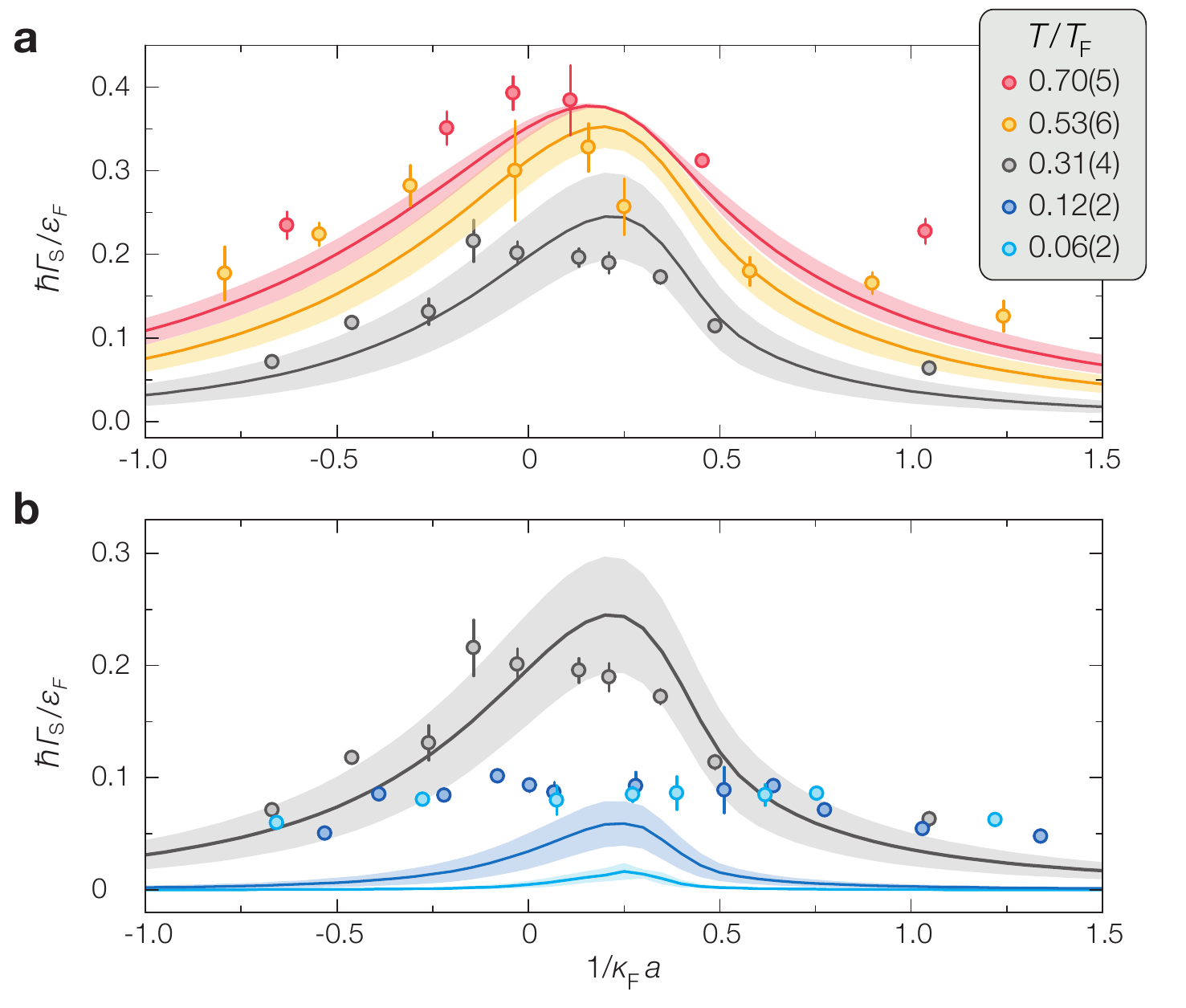}
\caption{\textbf{Spin drag coefficient of a strongly interacting Fermi gas.} $\hbar \Gamma_S/\epsilon_F$ is plotted as a function of $1/\kappa_F a$ for \textbf{a},  $0.31 \leq T/T_F \leq 0.7$, and \textbf{b}, $T/T_F \leq 0.31$. Experimental points are obtained by fitting the dynamics at $t>50\,$ms to the solution of a diffusion model (see Methods). Error bars combine uncertainties of the fit and of our determination of $\epsilon_F$. Lines are predictions from a T-matrix kinetic theory (see Supplementary Information), assuming the nominal initial $T/T_F$ and allowing a $\pm$20$\%$ temperature variation (shaded areas).}
\label{Fig4}
\end{center}
\end{figure}

\bigskip\noindent\textbf{Long-time diffusive dynamics and spin drag coefficient}\\Once spin diffusion is established \cite{Sommer2011, Enss2012, Bardon2014}, the analysis of the long-time evolution $\Delta M(t)$ (or equivalently $d(t)$) within a simple kinetic model (see Methods) allows us to determine also the spin drag coefficient $\Gamma_S$ as a function of temperature and interaction. The results are displayed in Fig.~\ref{Fig4}. These are compared with theoretical predictions for $\Gamma_S$, calculated for a single impurity diffusing in a homogeneous ideal Fermi gas within kinetic theory, accounting for scattering in all available states within $T$-matrix approximation for the scattering cross section (see Supplementary Information). 
The model is able to quantitatively reproduce the measured maximum of $\Gamma_S$ for $T/T_F \geq 0.3$, that is the expected range of validity of the $T$-matrix approximation, as well as the position of the maximum of $\Gamma_S$ at all temperatures (see Fig.~\ref{Fig4}). 
At low temperatures the data sets exhibit a small but appreciable asymmetry around the unitary point towards $\kappa_F a>0$.
Such a feature, which disappears progressively as temperature is increased, highlights the significant effect of collisions within the medium of surrounding particles on the dynamical properties of the diffusing quasi-particles (see Supplementary Information). A similar asymmetry in transport coefficients has already been reported for the shear viscosity \cite{Elliott2014} and transverse spin diffusion \cite{Trotzky2015}, but not in previous measurements of longitudinal spin diffusion \cite{Sommer2011}.

\smallskip
In conclusion, we have probed the ferromagnetic behaviour of the repulsive Fermi gas by investigating spin dynamics in a resonantly interacting ultracold \Li spin mixture. Our findings indicate that the short-time evolution of the system is governed by the effect of strong interparticle repulsion, and point to the existence of a Stoner-like ferromagnetic instability of the zero-range repulsive Fermi gas, before relaxation onto attractive states alters irrevocably the nature of the many-body state and its correlations. Furthermore, our study provides an important link between the dynamic response and the static properties of a strongly correlated fermionic state. In the future, our techniques may be exploited to probe spin transport and collective modes in different systems, such as repulsive and attractive Fermi gases with reduced dimensionality \cite{Levinsen2015}, and in the presence of weak optical lattices \cite{Pilati2014} and controlled disorder \cite{Pilati2016}.

\smallskip\noindent\textit{Note added} -- After submission of the manuscript, we became aware of related theoretical work\cite{He2016} by He~\textit{et al.}, reporting finite-temperature predictions qualitatively consistent with our findings.

\bigskip\bigskip\noindent {\bf METHODS}\\

\small
\noindent\textbf{Experimental protocols}\\
\noindent Our procedure to create weakly-interacting two-component Fermi mixtures of $^6$Li atoms has been already described elsewhere\cite{Burchianti2014} (see also Supplementary Information). The gas degeneracy parameter $T/T_F$ is adjusted by exploiting the tunability of the collisional properties of $^6$Li mixtures\cite{Zurn2013} during evaporation. We typically end up with $N_{\uparrow,\downarrow} \simeq 5 \times 10^4$\,atoms, confined in a cigar-shaped harmonic potential characterised by axial (radial) trap frequency of $\nu_z=21.0(1)$\,Hz ($\nu_{\bot}= 265(5)$ Hz).
To spatially separate the two spin components, at a magnetic field of about 1\,G, where the $\uparrow$ and $\downarrow$ states possess equal but opposite magnetic moments, we turn on a magnetic quadrupole gradient of about 1\,G/cm along the weak trap axis, that pushes the two spin clouds towards opposite directions. Once the overlap between the two components is zeroed, a repulsive optical potential centred at $z=0$, and characterised by a short (long) $1/e^2$ waist of $w_z= 2.0(2)$\,$\mu$m ($w_y=840(30)$\,$\mu$m) is employed to confine the two clouds into two disconnected reservoirs\cite{Valtolina2015}. These are characterised by an axial oscillation frequency $\nu_R=1.78(5)\,\nu_z$, and no appreciable particle tunnelling is detected over more than 2\,s with a barrier height of about 10\,$E_F$.

To excite the spin-dipole mode at a fixed Feshbach field, we abruptly switch off the barrier potential within less than 1\,$\mu$s. The two spin clouds, initially separated by a distance of about 5 $\mu$m, start moving one towards the other at a small relative velocity.
From here on, we monitor the dynamics of the relative distance between the centres of mass of the two components. For each value of evolution time, in two independent and successive experimental runs, we acquire two \textit{in-situ} absorption images with two high-intensity optical pulses, each of which is resonant only with one spin state.
For each point of interaction, temperature, and evolution time herein investigated, the relative distance between the two clouds is determined by averaging over at least five independent measurements.
To investigate the spin diffusion of the two domains into each other, and in particular to reveal the dynamical arrest of spin diffusion, we employ a different protocol turning off the repulsive barrier slowly in a two-stage sequence (see Supplementary Information for details). 

The centre of mass of each cloud is equivalently determined either via a Gaussian fit to the imaged density distribution (used for data displayed in Fig.~\ref{Fig2}a-c) or via the evaluation of the centre of mass $z_{i}$ of each sample $i=\uparrow,\downarrow$ obtained via direct integration of the bare images.
Once the spin dynamics is recorded, we fit $d(t)= (z_{\uparrow}- z_{\downarrow}) (t)$ to an exponential decay. 
By subtracting the fitted exponential drift from $d(t)$, we extract the spin-dipole collective dynamics $\Delta d(t)$, whose frequency is then derived by means of a damped sinusoidal function (see Supplementary Information).
Similarly, direct integration of the images at each evolution time allows to determine the relative population of the $i=\uparrow,\downarrow$ component in the left and right reservoirs, $M_{i}=(N_{i, L}-N_{i, R})/(N_{i,L}+N_{i,R})$, from which we obtain the magnetization $\Delta M= (M_{\uparrow}-M_{\downarrow})/2$. The behaviour of $d(t)$ closely reflects the one of $\Delta M(t)$.

\bigskip\noindent\textbf{Effective Fermi energy and wavevector}\\
\noindent We have defined $\kappa_F$ and $\epsilon_F$ as relevant length and energy scales. These are evaluated by approximating the thin barrier at $z=0$ as a delta-like potential. This is reasonable considering that the barrier thickness of 2\,$\mu$m is about 70 times smaller than the typical Thomas-Fermi radius of the cloud along the $z$-axis. Namely, we approximate the density distribution of a Fermi gas of $N$ atoms confined in a ``€œhalf''€ of the harmonic trap bisected by the thin barrier as half of the distribution of a Fermi gas of $2N$ atoms in the whole harmonic trap. The latter is evaluated using the finite-temperature Fermi-Dirac distribution of an ideal Fermi gas $n_F(\textbf{r},T/T_F)$, given the measured $T/T_F$, trap frequencies and atom number.
From $n_F(\textbf{r},T/T_F)$ the local wavevector and Fermi energy are given by $k_F(\textbf{r},T/T_F)=(6 \pi^2 n_F(\textbf{r},T/T_F))^{1/3}$ and $E_F(\textbf{r},T/T_F)= \hbar^2 k_F^2(\textbf{r},T/T_F)/(2 m_{\rm Li})$, respectively. 
The two parameters $\kappa_F$ and $\epsilon_F$ are derived by averaging $k_F(\textbf{r},T/T_F)$ and $E_F(\textbf{r},T/T_F)$ over a region as thick as the local interparticle spacing $(6 \pi^2)^{1/3}\!/ k_F(x,y,0,T/T_F)$ around $z=0$. Integration of $n_F(\textbf{r},T/T_F)$ within such region yields the total number $N_\text{int}$ of $\uparrow$ and $\downarrow$ fermions at the interface between the two spin domains (see Supplementary Information for details).

\bigskip\noindent\textbf{Sum-rule approach for the spin-dipole mode frequency}\\
\noindent Given a perturbation operator $D$, the moments of the strength distribution function, or sum rules, are given by $m_k=\sum_n |\langle 0| D |n\rangle|^2 (E_n-E_0)^k$. In particular, the spin-dipole mode is excited by the operator $D=\sum_{i\uparrow} z_{i} - \sum_{i\downarrow}z_{i}$, where $z$ is the longitudinal coordinate and $i_{\uparrow,\downarrow} = 1, \dots, N_{\uparrow,\downarrow}$. 
The frequency $\omega_{SD}$ of the spin-dipole oscillation is estimated using the ratio (see Supplementary Information and Ref. \citenum{Recati2010}):
\begin{equation}
\hbar^2\omega^2_{SD} = {m_1\over m_{-1}}\,.
\end{equation}
Within the local density approximation (LDA) one can obtain (see Supplementary Information)
\begin{equation} 
	\omega_{SD}^2=\frac{N}{m \int d\textbf{r} z^2 \chi(n)}\,,
\label{nuSDweighted}	
\end{equation}
where $\chi(n)$ is the zero-temperature magnetic susceptibility of a uniform gas obtained at the density $n$ from QMC calculations\cite{Pilati2010}.
The dashed green line plotted in Fig.~\ref{Fig2} is calculated through Eq.~\eqref{nuSDweighted} by inserting the density profile $n(r)$ of two fully overlapped spin components, determined in LDA using the equilibrium condition $\mu(n) - V_{\text{trap}} = \mu_0$, where $\mu(n)$ is the chemical potential as a function of local density $n$ from QMC calculations and $\mu_0$ is fixed by the normalisation condition.
The solid green line in Fig.~\ref{Fig2} accounts instead for a reduced overlap of 25$\%$ around the trap centre, in closer analogy to the experimental condition. In this case, the integral in Eq.~\eqref{nuSDweighted} over the outer spin-polarised regions contributes thus only with the spin susceptibility $\chi_0$ of an ideal Fermi gas, yielding a reduced deviation of the spin-dipole frequency from the bare trap frequency.
Therefore, we expect the measured frequencies to be higher than the results of Ref.~\citenum{Recati2010}, derived at full overlap. Consequently, the solid and dashed lines in Fig.~\ref{Fig2} delimit a confidence region in which most experimental data are found. Most importantly, the critical interaction strength at which the abrupt change in $\omega_{SD}$ occurs does not depend on the initial overlap configuration. 
Moreover, the spin diffusion dynamics happens on a longer timescale ($\gtrsim 200\,$ms), as displayed in the Supplementary Information, compared to the spin-dipole period ($\sim 50 - 100\,$ms). Hence, while we cannot assume that the system oscillates near an equilibrium configuration as in linear-response theory, the (slow) timescale for diffusion is sufficiently separated from the (faster) timescale for the spin oscillations. 

\bigskip\noindent\textbf{Polaron model for the domain-wall melting}\\
\noindent The proposed modelling of the plateau data shown in Fig.~\ref{Fig3} proceeds as follows. In the case of purely repulsive interaction, the ferromagnetic state, if energetically allowed, would be indefinitely stable and the miscibility of the two components would be prevented by the existence of a domain wall. In particular, a $\downarrow$ fermion at the interface would need to pay a finite amount of energy $\sigma>0$ in order to access the other spin domain forming a repulsive polaron at energy $E_+$. In our metastable system, however, if a repulsive polaron is created, it can subsequently decay onto the lower branch with a rate $\Gamma$, releasing an energy equal to the mismatch between the two branches, $E_+-E_-$. Hence, this two-step process will cause a net increase of energy $\Delta E= E_+-E_- - \sigma$ at a rate $\Gamma$. 
Importantly, the behaviour of $\Gamma$, $E_+$ and $E_-$ as a function of the interaction strength can be derived from recent non-perturbative theory approaches\cite{Schmidt2011, Massignan2014}.
We assume that at the beginning of the dynamics, the energy associated to the domain wall is given by $\sigma N_{\text{int}}$, $N_{\text{int}}$ being the total number of fermions within a slice around $z=0$ of total thickness equal to one interparticle spacing. The duration of the plateau $\tau_p$ is then set by the condition:
\begin{equation}
\sigma N_{\text{int}}= (E_+-E_- - \sigma) \tau_p \Gamma\,.
\label{Eq1}
\end{equation}
We assume that $\sigma= E_+ - E_{+c}$, where $E_{+}$ is the energy of a repulsive polaron, while $E_{+c}$ is the energy of one free fermion at the interface, left as a phenomenological parameter, and is independently adjusted for each $T/T_F$ herein investigated. From Eq.~\eqref{Eq1} we fit the experimental data by optimising the only free parameter $E_{+c}$ (see also Supplementary Information).

\bigskip\noindent\textbf{Diffusion model for extracting the spin drag coefficient}\\
\noindent The equation for the dynamics of the relative centre of mass $d=z_\uparrow - z_\downarrow$ can be easily obtained from the Boltzmann equation and is written as (see Supplementary Information)
\begin{equation}
 \ddot d+\Gamma_s \dot d+\omega_z^2 d=0,
 \label{eomd}
\end{equation}
where $\omega_z$ is the longitudinal trap frequency, and $\Gamma_s$ the spin drag coefficient due to collisions\cite{Sommer2011,Enss2012}.
We obtain the experimental spin drag coefficient by fitting the solution of Eq.~(\ref{eomd}) to the data, considering as initial condition $d(0)=d_0$
and ${\dot d}(0)=0$.

In Fig. ~\ref{Fig4} we compare the experimental results with a theoretical prediction based on a $T$-matrix approximation for the scattering cross-section corrected 
by the available scattering states in order to have a well defined scattering amplitude in the collisional integral (see Supplementary Information).
The agreement is especially good down to temperatures as low as $T=0.3\,T_F$.
At lower temperatures, both the shape and the magnitude of the spin drag coefficient as a function of the interaction compare more poorly.
This is expected since at very low temperature the T-matrix approximation is not quantitatively correct, and moreover the gas may suffer some heating during the dynamics due to decay processes, making its temperature higher than the one measured at the start of the dynamics, which is used for the comparison with the theory model.

\footnotesize
\noindent\textbf{Acknowledgments} -- We thank A. Morales and J. Seman for contributions in the early stage of the experiment, and G. Bertaina, G. M. Bruun, C. Di Castro, C. Fort, S. Giorgini, R. Grimm, W. Ketterle, P. Massignan, S. Pilati, R. Schmidt, W. Zwerger, M. Zwierlein and the LENS Quantum Gases group for many stimulating discussions. We thank H. Tajima and Y. Ohashi for providing us recent data of the lower branch spin susceptibility. This work was supported under European Research Council grants
no.~$307032$ QuFerm2D, and no.~$637738$ PoLiChroM. A.R. acknowledges support from the Alexander von Humboldt foundation. T.E. acknowledges the Physics Department, Sapienza University of Rome, for hospitality, and the Humboldt foundation for financial support during part of this work.


\smallskip\noindent\textbf{Additional information} -- Correspondence and requests for materials
should be addressed to M.Z.~(e-mail: zaccanti@lens.unifi.it).



\onecolumngrid
\begin{center}

\newpage
\vspace*{10mm}
\textbf{
\large \textsf{SUPPLEMENTARY INFORMATION}\\[4mm]
\large Exploring the ferromagnetic behaviour of a repulsive Fermi gas via spin dynamics}\\
\vspace{3mm}
\normalsize{G.~Valtolina,$^{1,2,3}$
F.~Scazza,$^{1,2}$
A.~Amico,$^{1,2}$
A.~Burchianti,$^{1,2}$\\
A.~Recati,$^{4,5}$
T.~Enss,$^{6}$
M.~Inguscio,$^{1,2}$
M.~Zaccanti$^{1,2,*}$
and~G.~Roati$^{1,2}$}\\
\vspace{0.5mm}
{\em \small
$^1$INO-CNR, Via Nello Carrara 1, 50019 Sesto Fiorentino, Italy\\
$^2$\mbox{LENS and Dipartimento di Fisica e Astronomia, Universit\`{a} di Firenze, Via Nello Carrara 1, 50019 Sesto Fiorentino, Italy}\\
$^3$Scuola Normale Superiore, Piazza dei Cavalieri 7, 56126 Pisa, Italy\\
$^4$INO-CNR BEC Center and Dipartimento di Fisica, Universit\`a di Trento, 38123 Povo, Italy\\
$^5$Technische Universit\"at M\"unchen, James-Franck-Stra{\ss}e 1, 85748 Garching, Germany\\
$^6$Universit\"at Heidelberg, Philosophenweg 19, 69120 Heidelberg, Germany\\}
{\small$^*$E-mail: \href{mailto:zaccanti@lens.unifi.it}{\small{zaccanti@lens.unifi.it}}}
\end{center}


\setcounter{equation}{0}
\setcounter{figure}{0}
\setcounter{table}{0}
\setcounter{section}{0}
\setcounter{page}{1}
\makeatletter
\renewcommand{\theequation}{S.\arabic{equation}}
\renewcommand{\thefigure}{S\arabic{figure}}
\renewcommand{\thetable}{S\arabic{table}}
\renewcommand{\thesection}{S.\arabic{section}}

\let\OLDthebibliography\thebibliography
\renewcommand\thebibliography[1]{
  \OLDthebibliography{#1}
  \setlength{\parskip}{3pt}
  \setlength{\itemsep}{3pt plus 0.3ex}
}


\normalsize
\vspace*{10mm}
\section{Experimental methods}
\label{Expcond}
	
\begin{figure}[t]
\centering
\includegraphics[width=0.9\columnwidth]{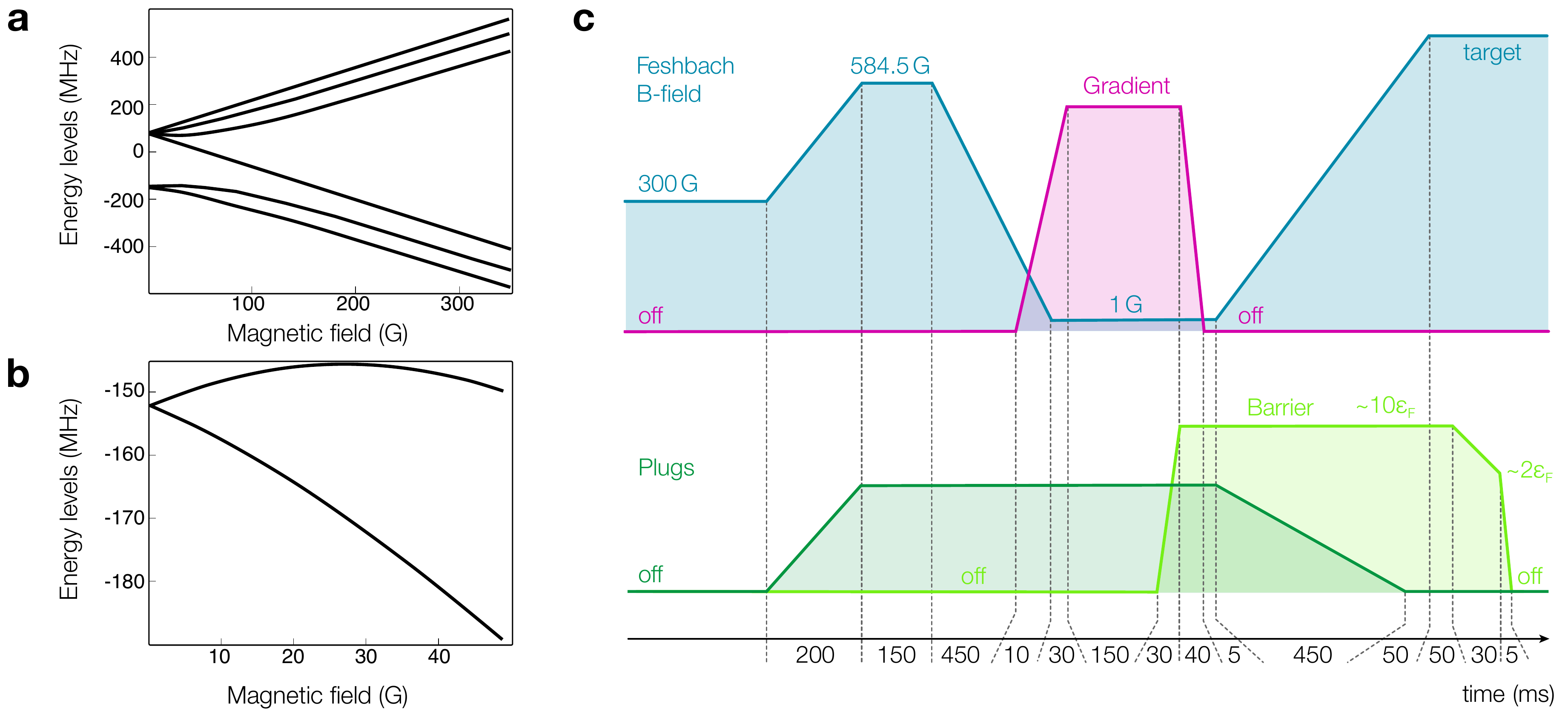} 
\caption{\textbf{a}, Hyperfine and Zeeman levels structure of $^6$Li in the $F=1/2$ and $F=3/2$ manifolds as a function of the magnetic field. \textbf{b}, Energy dependence of the two lowest Zeeman states ($|1\rangle$ and $|2\rangle$) at low magnetic field. Below 5\,G, the two states have magnetic moments with nearly equal amplitude but opposite sign. \textbf{c}, The experimental sequence used for spin separation is sketched: the Feshbach field (light blue), the magnetic gradient (purple), the side plugs (dark green) and the central barrier (light green) are plotted as a function of time.}
\label{procedure}
\end{figure}

\noindent\textbf{Fermi gas preparation procedure}
\newline	
Our procedure to create weakly interacting two-component Fermi mixtures of $^6$Li atoms has been already described in detail elsewhere \cite{Burchianti2014}.
To realize highly degenerate samples at $T/T_F < 0.2$, a balanced mixture of atoms equally populating the lowest and third-to-lowest Zeeman states (hereafter denoted $|1\rangle$ and $|3\rangle$, see Fig.~\ref{procedure}a) is evaporated in a crossed optical dipole trap (ODT) at a bias magnetic field of 300 Gauss. At such a field the $|1\rangle - |3\rangle$ mixture features a large \cite{Zurn2013}, though off-resonant, scattering length value of about -900 $a_0$ which makes the forced evaporation process extremely efficient \cite{Burchianti2014}. Here $a_0$ is the Bohr's radius.
We typically end up with $N_1 \simeq N_3 \simeq 5 \times 10^4$ atoms per spin component, confined in a cigar-shaped harmonic potential with axial and radial trap frequencies $\nu_z=21.0(1)\,$Hz and $\nu_{\bot}= 265(5)\,$Hz, respectively.

For the preparation of less degenerate samples, we either heat up in a controlled way the cloud by quickly turning off and on the ODT for an excitation time up to 1 ms, or we perform evaporation at 300 Gauss with a $|1\rangle - |2\rangle$ mixture. This features an interspecies scattering length of about -300$\,a_0$, which makes the evaporation process less efficient than for the $|1\rangle - |3\rangle$ spin combination, leading to similar atom number and trap frequencies, though at higher temperature.
The degeneracy parameter $T/T_F$ is determined by fitting the cloud density profiles \textit{in situ} or after $5\,$ms of time of flight to a finite temperature Fermi-Dirac distribution. 

At the end of evaporation in the $|1\rangle-|3\rangle$ mixture, we transfer all atoms from the $|3\rangle$ to the $|2\rangle$ state via a resonant 80$\,\mu$s radio-frequency (RF)  $\pi$-pulse. To avoid detrimental finite-state interaction effects, that would limit the transfer efficiency, we perform the transfer at a bias field of 584.5\,G, where the scattering lengths of the $|1\rangle-|3\rangle$ and $|1\rangle-|2\rangle$ spin mixtures are equal and non-resonant \cite{Zurn2013}. 

\bigskip
\noindent\textbf{Spin separation procedure}
\newline
The procedure for creating two separate spin domains is described here and summarized in Fig.~\ref{procedure}c.
In order to spatially separate the $|1\rangle$ and $|2\rangle$ components, we adiabatically lower the magnetic field down to $\sim 1\,$G, where the two states are essentially non-interacting and possess equal but opposite magnetic moments (see Fig.~\ref{procedure}b). 
Subsequently, we turn on a magnetic quadrupole gradient of about 1\,G/cm along the weak trap axis through a 40 ms linear ramp, shifting the two spin clouds in opposite directions. After 180 ms, once the overlap between the two components is completely zeroed, we turn on through a 40\,ms linear ramp a strongly anisotropic 532 nm optical beam \cite{Valtolina2015} centered at $z=0$, and characterized by a short (long) $1/e^2$ waist of $w_z=$2.0(2)\,$\mu$m ($w_x=$840(30)\,$\mu$m).
The beam, blue-detuned with respect to the 671\,nm lithium main transition, results in a repulsive potential which bisects the cloud along the weak axis into two reservoirs. The typical maximum barrier height after the 40 ms ramp is about 10 $E_F$, a value that impedes any appreciable particle tunnelling between the two reservoirs over more than 2\,s. Once the two spin clouds are separated and disconnected, we adiabatically turn off the magnetic gradient and increase the bias Feshbach field with a 500\,ms linear ramp up to the target value.
To prevent atom losses due to the magnetic field gradient, after the evaporation we turn on two additional elliptic plug beams at 532\,nm with short (long) waist of $w_z=50\,\mu$m ($w_y=120\,\mu$m). The short waist is aligned along the weak trap axis and the plugs enter perpendicularly to this, creating two effective repulsive walls that reduce as much as possible spilling of atoms due to the magnetic gradient. The plugs are turned on through a 200\,ms linear ramp after evaporation and switched off through a 450\,ms linear ramp when the Feshbach field is increased up to the final target value.
The overall spin-separation procedure causes a decrease of about 25$\%$ of the atom population, whereas no change of $T/T_F$ is measured. The degeneracy parameter after spin separation is estimated by a finite temperature Fermi-Dirac distribution fit of the density distribution of spin polarized clouds recorded after a 5\,ms time-of-flight expansion at a bias field of 300\,G. 

It is important to stress that the final configuration is perfectly symmetric, both in total spin population and density distribution, and each component does not present any appreciable shape nor center-of-mass oscillation within the separated reservoirs.
We note here that a measurement of the oscillation frequency of a single spin component within one isolated reservoir yields $\nu_R=1.78(5) \nu_z$. This is only about 10$\%$ lower than the value 2$\nu_z$ expected for the case of an infinitely thin barrier. This justifies the approximation of the barrier potential as delta-like for modelling the cloud density, as discussed in detail in Section~\ref{EFandkF}.

\bigskip
\noindent\textbf{Excitation and measurement of spin response}
\newline
In order to excite the spin-dipole mode at a fixed Feshbach field, we abruptly turn off the barrier potential from its maximum height of about $10\,E_F$ down to zero, within less than one $\mu$s. The two spin clouds, whose edges are initially separated by a distance of about 5\,$\mu$m, start moving one towards the other at a small, though non-zero, relative velocity.
From here on, we monitor the dynamics of the relative distance between the cloud centers of mass. For each value of the hold time, in two independent and successive experimental runs we acquire two \textit{in situ} absorption images, each of which is resonant only with one spin state.
For each value of interaction, temperature, and evolution time herein investigated, the relative distance between the two clouds is determined by averaging over at least five independent measurements.

\begin{figure}[b!]
\centering
\includegraphics[width=13cm]{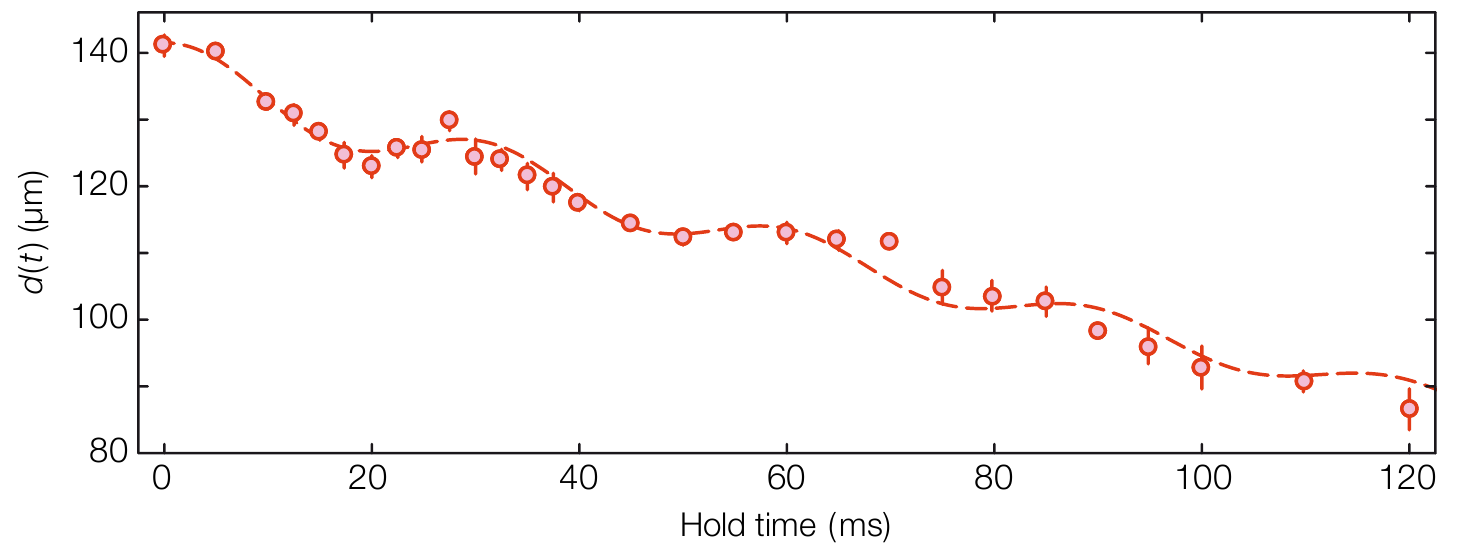}
\caption{Evolution of the center-of-mass distance between the two spin clouds, $d(t) = (z_{\uparrow} -  z_{\downarrow})(t)$, at $1/(\kappa_F a) \simeq 0.05$ and $T/T_F \simeq 0.12$. The dashed curve is obtained by fitting the data with the sum of an exponential decay and a damped sinusoidal oscillation.
}
\label{fig:d(t)}
\end{figure}

\newpage
The center of mass $z_{i}$ of the $i=\uparrow,\downarrow$ cloud is equivalently determined either via a Gaussian fit to the imaged density distribution (used for data displayed in Fig.~2a-c) or by directly evaluating $z_{i}$ via  integration of the bare images.
Once the spin dynamics is recorded (see e.g.~Fig.~\ref{fig:d(t)}), we fit $d(t)= ( z_{\uparrow} -  z_{\downarrow})(t)$  to an exponential decay. By subtracting the fitted exponential decrease of $d(t)$ due to diffusion from the data, we can isolate the spin-dipole mode, whose frequency is extracted from the data by means of a damped sinusoidal fitting function $f(t)= A \cos(2 \pi \nu_s\, t)\,e^{-t/\tau}$.
Similarly, direct integration of the images at each evolution time allows to determine the relative population of the $i=\uparrow,\downarrow$ component in the left and right reservoirs, $M_{i}=(N_{i, L}-N_{i, R})/(N_{i,L}+N_{i,R})$, from which we obtain the total magnetization $\Delta M= (M_{\uparrow}-M_{\downarrow})/2$. The behavior of $d(t)$ closely reflects the one of $\Delta M(t)$. 

\bigskip
\noindent\textbf{Measurement of magnetization evolution and spin diffusion}
\newline
As already mentioned in the main text, to investigate the spin diffusion of the two domains into each other, and in particular to reveal the dynamical arrest of spin diffusion, we employ a different protocol which does not excite the spin-dipole mode. For this purpose, we turn off the repulsive barrier slowly in a two-stage sequence. 
First, we lower the barrier from 10$\,E_F$ down to about $\leq 2\,E_F$ through a linear 30\,ms ramp (see Fig.~\ref{procedure}c). The ramp speed and final barrier height are chosen to ensure an adiabatic re-adjustment of the density distributions of the two clouds in the barrier region, while preventing a detectable tunnelling of atoms across the barrier and not inducing any shape excitations.
In this way, the magnetization in each reservoir does not change, but the relative distance between the edges of the two spin domains near $z=0$ is drastically reduced, from 5\,$\mu$m down to about 1\,$\mu$m, a length-scale comparable with the mean interparticle spacing of our gas at the interface.

\begin{figure}[h!]
\centering
\vspace*{10mm}
\includegraphics[width=\columnwidth]{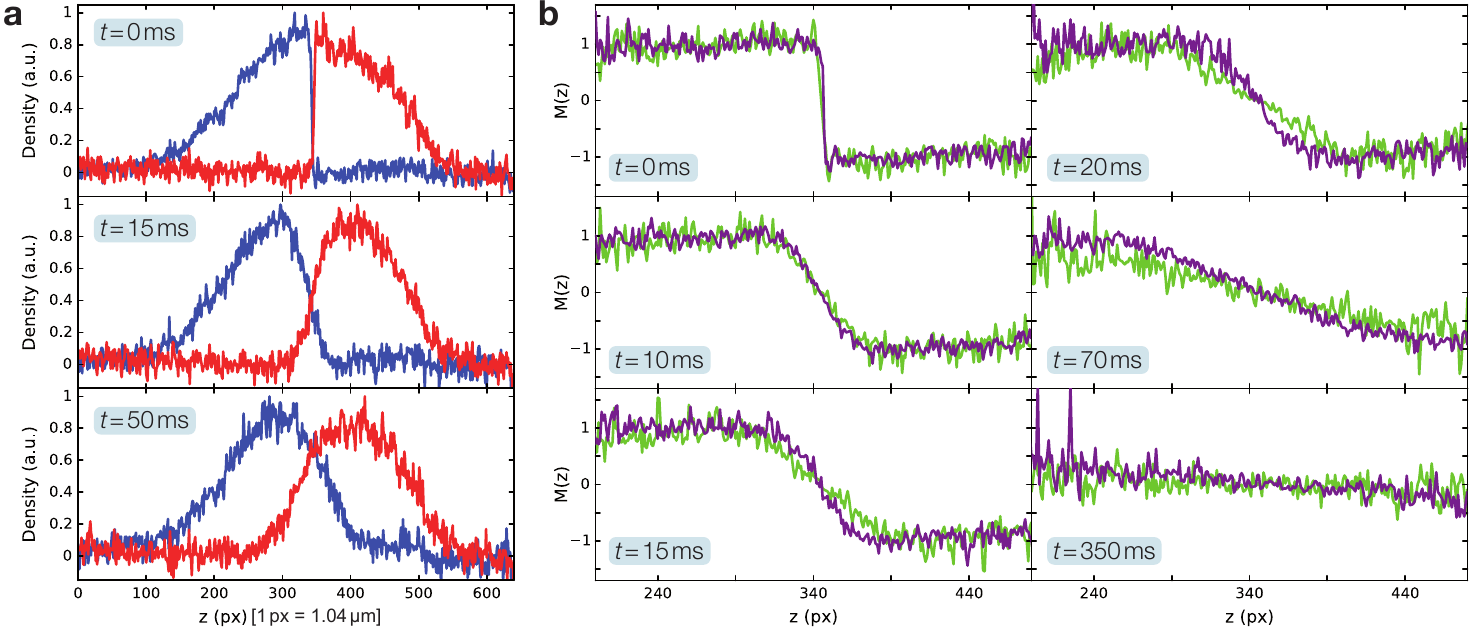}
\caption{\textbf{a}, Radially-integrated density profiles $n_{\uparrow}(z)$ (blue) and $n_\downarrow(z)$ (red) of the two spin domains at different hold times after slowly removing the optical barrier, with $1/(\kappa_F a) \simeq 0.05$ and $T/T_F \simeq 0.12$. \textbf{b}, Longitudinal magnetization $M(z)$ at different hold times for $1/(\kappa_F a) \simeq 0.05$ (purple) and $1/(\kappa_F a) \simeq 0.9$ (green) at $T/T_F \simeq 0.12$. It is visible here that $M(z)$ remains substantially unchanged for $1/(\kappa_F a) \simeq 0.05$ at hold times between 10 and 20\,ms.}
\label{fig:profiles}
\end{figure}

\begin{figure}[t!]
\centering
\includegraphics[width=0.6\columnwidth]{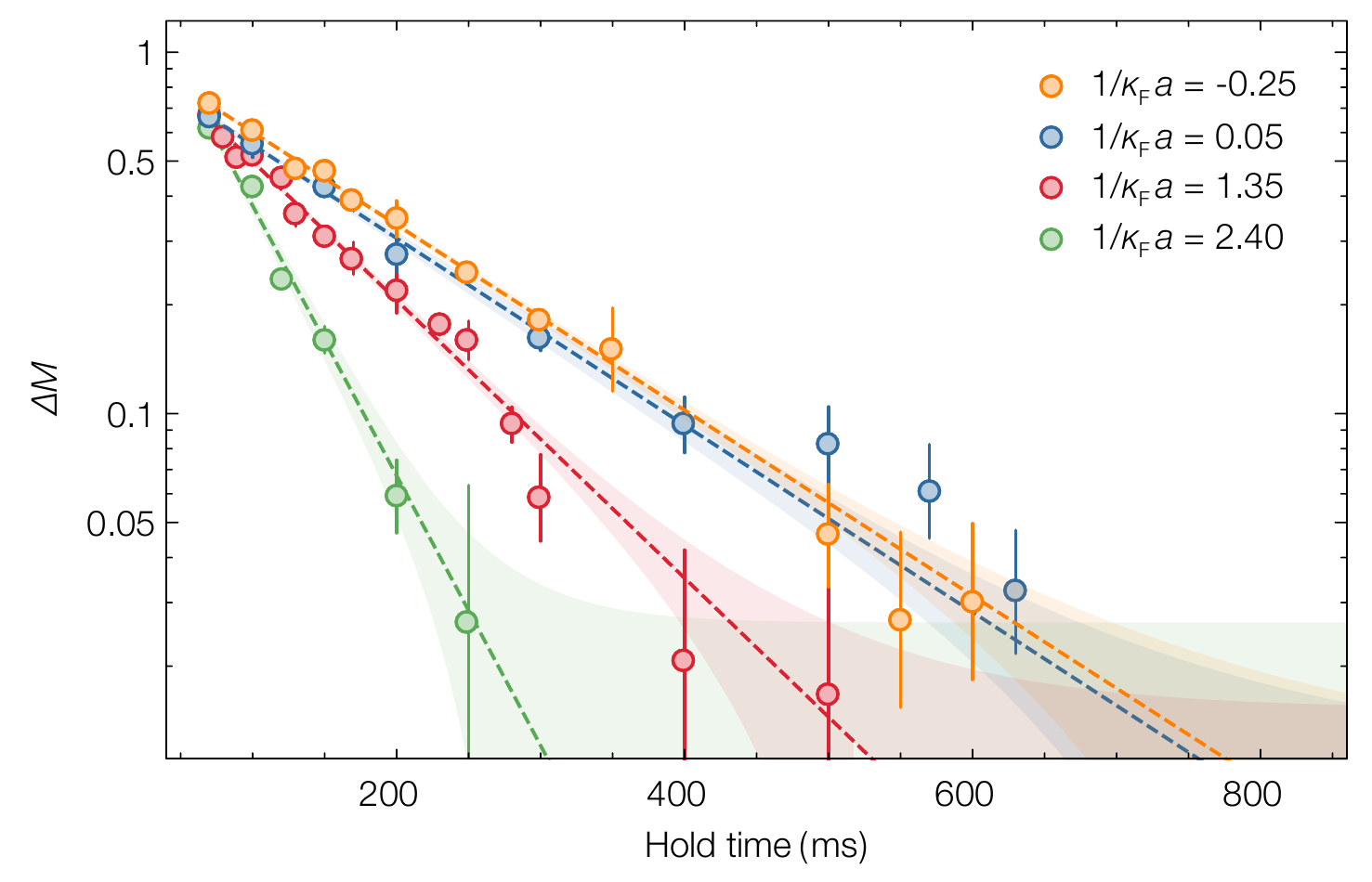} 
\caption{Spin diffusion dynamics at hold times $t > 50\,$ms for $T/T_F = 0.12(2)$. The evolution of the magnetization $\Delta M(t)$ is plotted for various interaction strengths. Data points averaged over 4-5 indipendent experimental realizations (with error bars representing the s.e.m.\! of the data) are shown together with fits to the solution of the simple kinetic model from which we extract $\Gamma_S$ (see Eq.~\eqref{diffeqn}).}
\label{diffusion}
\end{figure}

In a second stage, we ramp the intensity of the repulsive barrier down to zero.
We have investigated different durations of the second ramp, spanning from 0 up to 30\,ms.  
For ramps durations between 10 and 30\,ms, we detect an appreciable flow of atoms across the barrier region already during the ramp. In this case, once the barrier is off, the spin dynamics is well described by a continuous, single exponential decay.
For ramps shorter than 10\,ms, in the strongly interacting regime, we observe a $\tau_p > 0$ window of vanishing spin diffusion. 
For ramp times below 5\,ms, the duration of the plateaus is maximized and, for each target field explored in this work, it does not show any dependence on the ramp speed. 
For the measurements presented in Fig.~3 and 4 of the main text, the barrier is turned off in 5\,ms. Examples of the short-time evolution of the cloud density profiles and of the longitudinal magnetization $M(z) = (n_\uparrow - n_\downarrow)(z)/(n_\uparrow + n_\downarrow)(z)$ are displayed in Fig.~\ref{fig:profiles}. Consistently with the trend of the total magnetization $\Delta M$ shown in Fig.~3 of the main text, the evolution of the longitudinal magnetization at strong interactions exhibits a halt after a short evolution time (see Fig.~\ref{fig:profiles}b). 
After spin diffusion has been established at $t>50$\,ms, the evolution of the magnetization at each temperature and interaction strength is well described by Eq.~\eqref{diffeqn} (see e.g. Fig.~\ref{diffusion}).

\newpage

\newpage
\section{Effective Fermi energy and Fermi wavevector}
\label{EFandkF}
\begin{figure}[h!]
\centering
\vspace*{3mm}
\includegraphics[width=0.8\textwidth]{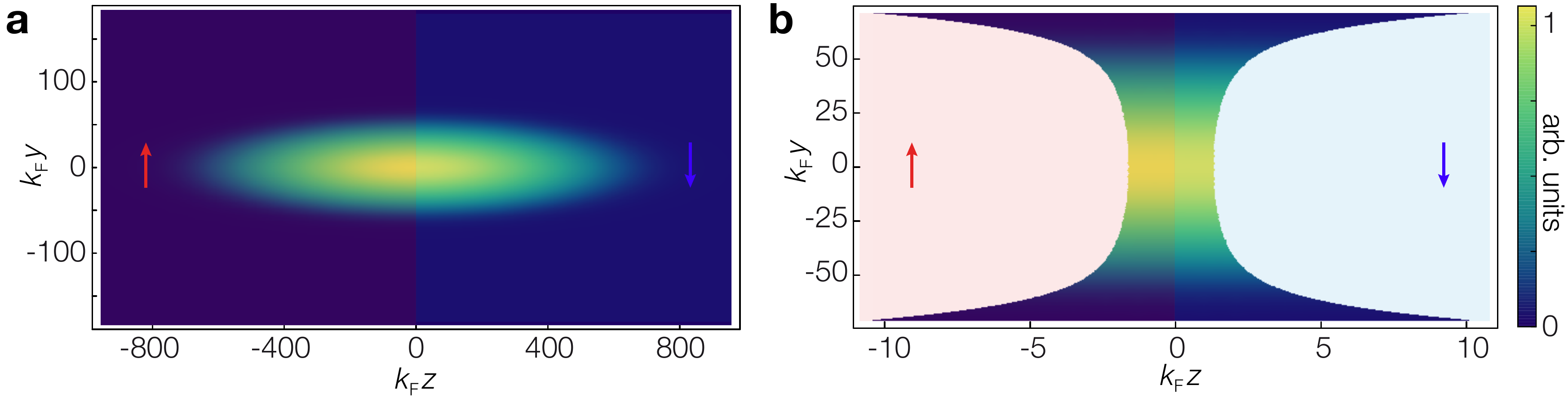}      
\caption{\textbf{a}, Model for total density distribution of the initial state: the optical barrier is approximated as infinitely thin, and the total profile of the two spin components is given by the one of a Fermi gas of $N_{\uparrow}+N_{\downarrow}$ atoms at the given temperature, occupying the whole trap. \textbf{b}, The values of the effective Fermi energy $\epsilon_F$ and wave vector $\kappa_F$ are obtained by averaging over a slice (shown) around $z=0$ of total thickness equal to one local interparticle spacing.}
\label{EffectiveEFandkF}
\end{figure}

In order to account for the inhomogeneity, the finite temperature, and the initial density of the two spin components in disconnected reservoirs, in the main text we have defined $\kappa_F$ and $\epsilon_F$ as relevant length and energy scales. 
These are evaluated by approximating the thin barrier at $z=0$ as a delta-like potential. As already discussed in Section 
\ref{Expcond}, this is a reasonable assumption, considering that the barrier thickness of 2\,$\mu$m is about 70 times smaller than the typical Thomas-Fermi radius of the cloud along the $z$ axis. This is also confirmed by the measurement of the oscillation frequency within each reservoir, that is found to be $\sim10\%$ lower than the value $2 \nu_z$ expected for the case of a delta-like potential. Namely, we approximate the density distribution of a Fermi gas of $N$ atoms confined in ``half'' a harmonic trap as half of the distribution of a Fermi gas of 2$N$ atoms occupying the whole dipole trap. The latter is evaluated using the finite-temperature Fermi-Dirac distribution of an ideal Fermi gas $n_F(\textbf{r},T/T_F)$, given the measured $T/T_F$, trap frequencies and atom number (see Fig.~\ref{EffectiveEFandkF}a). The obtained profile excellently reproduces the two $\uparrow$ and $\downarrow$ clouds, imaged \textit{in situ} independently.
From $n_F(\textbf{r},T/T_F)$ the local Fermi energy and wavevector are given respectively by $k_F(\textbf{r},T/T_F)=(6 \pi^2 n_F(\textbf{r},T/T_F))^{1/3}$ and $E_F(\textbf{r},T/T_F)= \hbar^2 k_F(\textbf{r},T/T_F)^2/(2 m_{\rm Li})$. 

The two parameters $\kappa_F$ and $\epsilon_F$ are derived by averaging the local $k_F(\textbf{r},T/T_F)$ and $E_F(\textbf{r},T/T_F)$ over a region as thick as the local interparticle spacing $(6 \pi^2)^{1/3}\!/k_F(x,y,0,T/T_F)$ around $z=0$ (see Fig.~\ref{EffectiveEFandkF}b). Parallel to this, integration of $n_F(\textbf{r},T/T_F)$ within this region yields the total number $N_{int}$ of $\uparrow$ and $\downarrow$ fermions at the interface between the two spin domains.
The definition of such an energy and length scale are inspired by a recent study on the spatial distribution of a fully ferromagnetic two-fermion mixture in a harmonic trap \cite{Zintchenko2013}, which predicts that the domain wall at the interface between the two magnetic domains has a total thickness of one interparticle spacing.

\section{Sum-rule approach for the prediction of the spin-dipole frequency}
\label{SD_Freq}
The spin-dipole mode is an oscillation of the relative position of the center of mass (COM) of each cloud of $\ket{\uparrow}$ and $\ket{\downarrow}$-atoms, but not of their total COM position, i.e. a motion where atoms of one species move in the opposite direction with respect to the atoms of the other species. For this reason, it is also referred to as an out-of-phase (spin) mode.
As discussed in the main text, the measurement of the spin-dipole mode frequency provides a spectroscopic tool for disclosing the spin susceptibility of the gas. An intuitive reasoning can clarify the relation between the mode frequency and the susceptibility, expressed in Eq.~(1) of the main text.
A finite frequency of the spin-dipole mode is due to the presence of a restoring force when one tries to pull apart $\ket{\uparrow}$ and $\ket{\downarrow}$-particles in opposite directions. With no interactions, the only restoring force is provided by the trapping potential and the COM of each cloud oscillates at the trap frequency. A repulsive (attractive) interaction between the two atomic states leads to a smaller (larger) energy cost to pull them apart with respect to the non-interacting case. This leads to a smaller (larger) restoring force in the equation of motion, eventually resulting in a frequency smaller (larger) than the trap frequency. We indeed observe a reduced (increased) frequency of the spin dipole mode when the system is prepared on the repulsive (attractive) branch (see Fig.~\ref{attr-spin-dipole}). For similar reasons, the spin susceptibility of a Fermi gas increases (decreases) if the interspecies particle interaction is repulsive (attractive). Therefore, a frequency decrease of the spin dipole corresponds to an increase of the spin susceptibility, and vice versa.
The previous argument can be made more formal using a sum-rule approach \cite{sumrules, Recati2010}, which provides an upper bound of the spin-dipole mode frequency in terms of the system susceptibility. When the response is characterised by a single delta-function (in the dynamic structure factor), the upper bound becomes an exact result.
As emphasized in the main text, the measured spin-dipole oscillation frequency compares well with the sum-rule estimate supplemented with quantum Monte Carlo (QMC) data and employing a local density approximation \cite{Recati2010}. 

The sum-rule approach is based on the knowledge of different momenta of the linear response of the system to the perturbation exciting the mode of interest. 
Given a perturbation operator $D$, the moments of the strength distribution function, or sum rules, are given by $m_k=\sum_n |\langle 0| D |n\rangle|^2 (E_n-E_0)^k$. An upper bound to the mode frequency is provided by the ratios of the sum rules.
In particular, the spin-dipole mode is excited by the operator $D=\sum_{i\uparrow} z_{i} - \sum_{i\downarrow}z_{i}$, where $z$ is the longitudinal coordinate and $i_{\uparrow,\downarrow} = 1, \dots, N_{\uparrow,\downarrow}$. 
We estimate the frequency of this mode using the ratio \cite{Recati2010}:
\begin{equation}
\hbar^2\omega^2_{SD} = {m_1\over m_{-1}}
\label{1-1}
\end{equation}
The reason is two-fold: (i) the inverse energy-weighted sum rule $m_{-1}$ is sensitive to low frequencies, which dominate since the stronger the repulsive interaction the softer the spin-dipole mode; (ii) $m_{-1}$ is related to the susceptibility of the system. \\
The so-called energy weighted sum rule $m_1$ is easily calculated in terms of a double commutator:
\begin{equation}
m_1= \frac{1}{2}\langle 0 | \left[D,\left[H,D\right]\right]|0\rangle = N{\hbar^2\over 2 m}.
\label{f}
\end{equation} 
The inverse energy-weighted sum rule is proportional to the spin dipole susceptibility. The latter can be determined by minimizing the total energy of the system in the presence of an external static coupling of the form $-\lambda D$. \\
Since our system is inhomogeneous, we can write the energy within the local density approximation (LDA):
\begin{equation}
E= \int d{\bf r}[\epsilon(n_\uparrow({\bf r}),n_\downarrow({\bf r})) -\lambda z(n_\uparrow({\bf r})-n_\downarrow({\bf r}))]
\label{E}
\end{equation}
where $\epsilon(n_\uparrow,n_\downarrow)$ is the energy density of the uniform gas. By expanding $\epsilon(n_\uparrow,n_\downarrow)$ up to quadratic terms in $n_\uparrow-n_\downarrow$, minimization of $E$ yields the result $n_\uparrow-n_\downarrow= \lambda z \chi(n)$ for the polarization induced by the external field. 
Here $\chi^{-1}(n)=\partial^2 \epsilon/\partial (n_\uparrow-n_\downarrow)^2$ is the zero-temperature inverse magnetic susceptibility of a uniform gas.\\ 
The calculation of the induced spin-dipole moment then yields $m_{-1}= 1/2\int d{\bf r} z^2 \chi(n)$ and hence the result (Eq.~(1) in the main text)
\begin{equation}
\omega^2_{SD}=\frac{N_{\uparrow}+N_{\downarrow}}{m \int d\textbf{r} z^2 \chi(n)}  
\label{chichi0}
\end{equation}
for the spin dipole frequency. Eq.~\eqref{chichi0} is exact within LDA and it shows that an increase of the magnetic susceptibility will result in a decrease of $\omega_{SD}$. If a ferromagnetic transition occurs, the spin-dipole mode frequency exhibits a minimum associated with a diverging spin susceptibility within the central high-density region of the trap \cite{Recati2010}. Once a ferromagnetic domain is formed, Eq.~\eqref{chichi0} is no longer valid and the dynamical behaviour of the system strongly depends on the geometry of the domain wall itself. 

For an ideal Fermi gas trapped in a harmonic potential one gets the simple result $\omega_{SD}=\omega_z$  where $\omega_z$ is the trap frequency along the $z$-axis.
For a balanced interacting Fermi gas on the repulsive branch, both for the energy density and the bulk magnetic susceptibility, we use the QMC results by Pilati {\sl et al.} \cite{Pilati2010}. Both quantities can be written easily as an expansion in $k_F a$ above the mean-field results:
\begin{eqnarray}
\frac{\epsilon}{\epsilon_0}&=&1+ k_Fa + C_\epsilon(k_F a)^2\\
\frac{\chi_0}{\chi}&=&1-\frac{2}{\pi}k_Fa-C_\chi(k_Fa)^2,
\label{MCfit}\end{eqnarray}
where $\epsilon_0$ and $\chi_0$ are the chemical potential and the susceptibility of a spin-$1/2$ free Fermi gas, respectively. The constants $C_\epsilon=0.28$ and $C_\chi=0.62$ are obtained by fitting to the QMC  results \cite{Pilati2010}.
The previous expressions are suitable to determine an upper bound to the spin-dipole mode frequency for an unpolarized gas at equilibrium in the trap \cite{Recati2010}. 

\enlargethispage{\baselineskip}
Providing a rigorous theoretical description of the experimental results shown in Fig.~2 of the main text is challenging since the gas has a time- and space-dependent local polarization.
The dashed green line in Fig.~2 is calculated through Eq.~\eqref{chichi0} by assuming a cylindrically symmetric density profile $n(r)$ with fully overlapped spin components, determined in LDA using the equilibrium condition $\mu(n) - V_{\text{trap}} = \mu_0$, where $\mu(n)$ is the chemical potential as a function of local density $n$ from QMC calculations and $\mu_0$ is fixed by the normalization condition.
The solid green line in Fig.~2 accounts instead for a reduced overlap of 25$\%$ around the trap center, in closer analogy to the experimental condition. 
In this case, the density profile used to compute Eq.~\eqref{chichi0} is comprised of three longitudinal regions: two external non-interacting polarized tails and a central part where the two components fully overlap. For small oscillations, the overlap region can still be described through $\mu(n)$, i.e. the equation of state of the repulsive branch. The size of the overlap region is chosen according to the typical experimental one. In particular, a value of 25\% denotes a ratio of 0.25 between the axial extension of the overlap region and the total axial size of the system.
The integral in Eq.~\eqref{chichi0} over the outer spin-polarized regions contributes thus only with the spin susceptibility $\chi_0$ of an ideal Fermi gas, yielding a reduced deviation of the spin-dipole frequency from the bare trap frequency.
Therefore, we expect the measured frequencies to be higher than the results of Ref.~\citenum{Recati2010}, derived at full overlap. The solid and dashed lines in Fig.~2 delimit a confidence region in which most experimental data are found. Most importantly, the critical interaction strength at which the abrupt change in $\nu_{SD}$ occurs does not depend on the initial overlap configuration. 
Moreover, the spin diffusion dynamics happens on a longer timescale ($\gtrsim 200\,$ms), as displayed in Fig.~\ref{diffusion}, compared to the spin-dipole period ($\sim 50 - 100\,$ms). Hence, while we cannot assume that the system oscillates near an equilibrium configuration as in linear-response theory, the (slow) timescale for diffusion is sufficiently separated from the (faster) time scale of the spin oscillations.

\section{Temperature shift of the critical interaction strength}

Within Landau-Fermi liquid theory the inverse susceptibility at $T=0$ can then be written as \cite{PinesNozieres}
\begin{equation}
\chi^{-1}(T=0)=\frac{2}{g(e_F)}(1+F_0^a)
\end{equation}
where $F_0^a$ is the $l=0$ antisymmetric (magnetic) Landau parameter, $g(e_F)=m^*k_F/\pi^2$ is the density of states  with $m^*=m(1+F_1^s/3)>m$ and
$F_1^s$ is the $l=1$ symmetric (density) Landau parameter. To the first order in the interaction one has $m^*=m$ and $F_0^a=-2k_F a/\pi$.

At finite temperature but $T\ll T_F$ we can consider the corrections to previous expression only due to free quasiparticles which is proportional 
to $n$, and therefore to $T^2$. The interaction term depending on $n^2$ will contribute to higher order $O(T^4)$:
\begin{equation}
\chi^{-1}(T)=\frac{2}{g(e_F)} \left(1+F_0^a+{\pi^2 \over 12} {T^2 \over T_F^2}\right)
\end{equation}

The paramagnetic state becomes unstable when $\chi^{-1}=0$, which at zero temperature can occur if there exists a critical value 
$(k_Fa)_c$ such that $F_0^a=-1$.
At low $T$ we can expand the Landau parameters around their values at the critical point $(k_Fa)_c$ at zero temperature 
\begin{eqnarray}
F_0^a&=&F_{0,c}^a+\left(\frac {\partial F_0^a}{\partial (k_F a)}\right)_c(k_Fa-(k_Fa)_c)\\
F_1^s&=&F_{1,c}^s+\left(\frac {\partial F_1^s}{\partial (k_F a)}\right)_c(k_Fa-(k_Fa)_c).
\end{eqnarray}
Therefore we find that the critical temperature for the paramagnetic state to be unstable can be simply written as
\begin{equation}
\left(\frac{T}{T_F}\right)_c=\frac{2\sqrt{3}}{\pi} \sqrt{-\left(\frac {\partial F_0^a}{\partial (k_F a)}\right)_c}\sqrt{k_Fa-(k_Fa)_c}
\label{power}
\end{equation}
For instance using the expression for $F_0^a$ at the second order in $k_F a$ \cite{Duine2005} one gets the first correction to the usual expression. In particular 
$(k_Fa)_c\simeq 1.05$ and the critical temperature reads
\begin{equation}
\left(\frac{T}{T_F}\right)_{\!\!c} \simeq\frac{2^{3/2}\sqrt{3}}{\pi^{3/2}}\sqrt{1+(k_Fa)_c\frac{8}{3\pi}(1-\ln 2)}\,\sqrt{k_Fa-(k_Fa)_c}
\end{equation}

In principle, application of the aforementioned power law as a fit to the experimental data in Fig.~3d is not justified: in fact, $\tau_p>0$ is interpreted as the region of metastability of the fully ferromagnetic state, whereas Eq.~\eqref{power} marks the boundary between paramagnetic and partially ferromagnetic phases. 
On the other hand, however, all QMC results show that the partially ferromagnetic phase for a homogeneous system occupies a very narrow region of the phase diagram, and we expect that in our trapped system its presence is essentially washed out by the inhomogeneity of the clouds density distribution and finite experimental resolution.

\bigskip
\section{Pairing instability and molecule formation during spin dynamics}
One major issue that has hindered the study of strongly repulsive Fermi gases in previous experiments \cite{Jo2009,Sanner2012} is represented by the pairing instability \cite{Pekker2011}. As discussed in the main text, ferromagnetic behaviour develops along the upper branch of the many-body system: however, this state features an additional instability, represented by the tendency of the paramagnetic phase to turn via inelastic processes into a gas of pairs, which represents the true ground state of the balanced system at low temperatures. At least for homonuclear mixtures and broad resonances \cite{Pekker2011,Sanner2012,Massignan2014}, the pairing instability always overcomes the Stoner's one.

Thanks to our preparation scheme, that artificially initialises the system into a fully ferromagnetic configuration, we are able to contain the system tendency towards pairing, allowing for the investigation of the metastable upper branch. Furthermore, as discussed below in Section~\ref{FM_melt}, attractive polarons, rather than pairs, seem to be the preferential decay products in our system, at least in the strongly interacting regime. Nonetheless, molecule formation has represented a major issue in previous experiments, and ruling out pairing effects for explaining the dynamics observed in our studies is fundamental for further supporting our interpretation.

In previous experiments \cite{Sanner2012}, the population of molecules and atoms has been identified via a rapid magnetic field ramp technique. After some evolution time at a target field close to the Feshbach resonance center, where molecules could be formed via recombination processes, two successive absorption images were acquired. A first, taken at high field, allowed to monitor the total population of atoms and molecules, since the molecule binding energy close to the Feshbach resonance is two to three orders of magnitude smaller than the natural linewidth of the imaging transition. This allows to take a picture of the molecules with the same imaging light employed for the atom, being the latter in the $\uparrow$ or $\downarrow$ state.
A second imaging pulse was taken after a rapid sweep of the magnetic field to zero. The ramp converts weakly bound pairs into deeply bound molecules, which become transparent to the atom imaging light. Hence, the second imaging selectively monitors the atom population, i.e. the population of the upper branch.

Our setup does not allow to perform fast ($\sim$10$^2\,$G$/\mu$s) ramps to low fields, hence we employed a different protocol to monitor the presence of molecules in the system. This is based on acquiring, within a single experimental run, two subsequent absorption images, by means of 4\,$\mu$s long pulses resonant with the $\uparrow$ and $\downarrow$ states, respectively, and separated by 300\,$\mu$s.
If no molecules are present, the effect of the first imaging pulse resonant with the $\uparrow$ state on $\downarrow$ atoms is found to be negligible, since the $\uparrow$ imaging light is off-resonant to the $\downarrow$ component. Furthermore, the short delay between the two pulses greatly limits the effect of heating of the $\downarrow$ cloud associated to collisions with escaping $\uparrow$ imaged atoms.
The effect of the first imaging pulse is completely different if molecules are present. Since the $\uparrow-\downarrow$ dimers are only weakly bound, the first imaging photon, resonant with the $\uparrow$ optical transition, dissociates the bound state into two atomic products \cite{Chin2005}, each of which symmetrically acquires a significant momentum. The latter is associated to the density of states of the two outgoing atoms, to the binding energy of the dimer (negligible in this case), and to the photon momentum $\hbar k_{L}$. The increase of the cloud size detected by the second imaging pulse is thus directly related to the amount of molecules in the system. We therefore monitor the increase of the radial width after the first pulse for different interaction strengths and different evolution times during the spin diffusion.

\begin{figure}[b!]
\centering
\includegraphics[width=10cm]{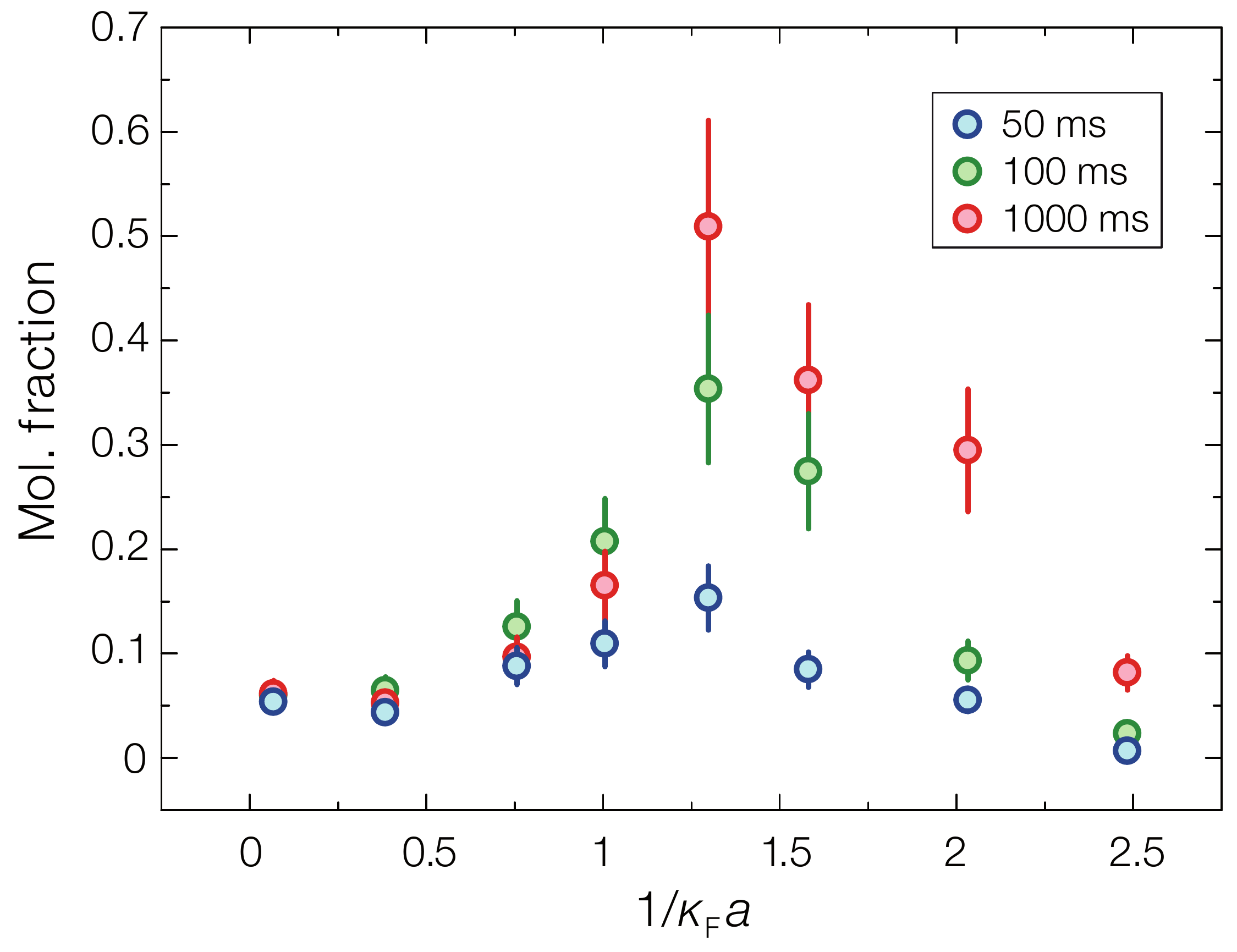}
\caption{Measured molecular fraction at $T/T_F = 0.12(2)$ for different evolution times after the barrier removal, obtained through the double-pulse imaging technique. The peak of molecule formation appears below the measured critical interactions, at $1/(\kappa_F a) \simeq 1.3$.}
\label{double_imaging}
\end{figure}

A simple model allows to link the molecular fraction to the increase of the cloud width after the first pulse, following the experimental protocol described above.  
In general, the size of a trapped cloud can be written as:
\begin{equation}\label{mol_fra1}
\langle x_0^2\rangle = \frac{2\langle U\rangle}{m\omega^2}
\end{equation}
Where $\langle U\rangle$ is the potential energy of one atom weighted over the density distribution of the cloud, the latter being eventually modified by interaction effects. 
In the case of a pure gas of dimers, application of an imaging pulse resonant with the $\uparrow$-component causes the dimers to dissociate with a certain transfer of energy $E_1$ to the $\downarrow$-component. Since the photon energy is always larger than the binding one, we assume $E_1$ to be independent from the molecular binding, and hence independent from $\kappa_Fa$. 
According to the same argument of Eq.~\eqref{mol_fra1}, the width measured through the second pulse, following the first, can be written as:
\begin{equation}\label{mol_fra2}
\langle x_1^2\rangle = \frac{2\langle U+E_1\rangle}{m\omega^2}
\end{equation}
If the gas is a mixture of $N_a$ free atoms and $N_m$ molecules, the mean size is set by:
\begin{equation}\label{mol_fra3}
\langle x^2\rangle = \frac{N_a\langle x_0^2\rangle +N_m\langle x_1^2\rangle }{N_a+N_m}
\end{equation}
Defining the molecular fraction of the gas at time $t$ during the spin dynamics, $f_m \equiv f_m (\kappa_F a, t) = N_m/N$, we get:
\begin{equation}\label{mol_fra4}
\langle x^2\rangle =\frac{2}{m\omega^2}\langle U\rangle +\frac{2}{m\omega^2}\langle E_1\rangle f_m =\langle x_0^2\rangle +\frac{2}{m\omega^2}\langle E_1\rangle f_m
\end{equation}
When starting from a pure molecular sample, $f_m=1$, we would have:
\begin{equation}\label{mol_fra5}
\frac{2}{m\omega^2}\langle E_1\rangle =\langle x^2_{1m}\rangle - \langle x^2_{0m}\rangle
\end{equation}
We can therefore express the molecular fraction as:
\begin{equation}\label{mol_fra5}
f_m =\frac{\langle x^2_{1}\rangle - \langle x^2_{0}\rangle }{\langle x^2_{1m}\rangle - \langle x^2_{0m}\rangle}
\end{equation}
For every investigated interaction strength, the denominator of Eq.~\eqref{mol_fra5} is experimentally determined by applying the double-pulse imaging technique on a superfluid gas with the same interaction parameter $\kappa_F a$ and a temperature $T/T_F<0.1$, ensuring a molecular fraction $f_m\simeq1$ on the BEC side of the resonance. The change of the density distribution when moving from the unitary limit to the BEC one is accounted by renormalizing the measured radial width by the average density of the gas, evaluated with the conventional single-pulse absorption imaging.\\
The numerator of Eq.~\eqref{mol_fra5} is evaluated by measuring the second-pulse radial size of the cloud $\langle x^2_{1}\rangle$ at different evolution times during the spin dynamics, initialized by the same experimental procedure discussed in Section~\ref{Expcond} and in the main text. $\langle x^2_{0}\rangle$ is in turn the size measured through the first imaging pulse at the corresponding times. 
Results of this analysis are reported in Fig.~\ref{double_imaging}, for various interaction strengths and different evolution times after abruptly removing the barrier, for a repulsive Fermi gas mixture initially prepared at $T/T_F=0.12(2)$.

The general trend for these measurements is interpreted as follows. In the weakly repulsive regime, the upper branch is very long-lived, and despite rapid mixing of the two spin clouds only a small number of molecules is formed. Increasing interactions, the decay rate of the upper branch monotonically increases \cite{Schmidt2011,Massignan2011,Kohstall2012,Massignan2014}: hence, despite an increase of the spin drag coefficient \cite{Duine2010,Sommer2011}, which tends to slow down the diffusion and reduces the spatial overlap of the $\uparrow$ and $\downarrow$ clouds, the molecule formation becomes more sizable, reaching a maximum near $1/(\kappa_F a) \simeq 1.3$, in good agreement with Ref.~\citenum{Zintchenko2013}. 
However, as one accesses the strongly interacting regime where ferromagnetism is promoted, the molecule formation is again strongly reduced, highlighting the tendency of the system to suppress the overlap between the two spin components.
The trend at large $\kappa_F a$ values, which persists also after long evolution times when the two clouds have come together, suggests that also at small values of local population imbalance the system may be a Fermi liquid state of attractive polarons, rather than a Bose gas of dimers. The Fermi liquid state might be favored by our way of initializing the system dynamics, as well as by the temperature increase associated with the exothermic decay process from the upper to the lower branch. 
Importantly, for timescales below 100\,ms, over which both the magnetization plateau and the spin-dipole oscillations were measured, the observed heating is relatively small and the derived molecule fraction remains below 10$\%$ for interaction strengths exceeding the critical value for the arrest of spin diffusion to occur.
We therefore conclude that neither the behavior of the spin-dipole frequency nor the appearance of plateaus in the spin diffusion can arise from dimer formation. 

This conclusion is further supported by the measured temperature dependence of the critical repulsion strength (see Fig.~3d and Eq.~\eqref{power}). Such a trend is antithetic to the one that would result if the observed spin dynamics was governed by the molecular population within the overlap region. Assuming that such a molecular layer is in local chemical equilibrium with the atoms owing to fast thermalization, the molecular fraction is set by the Boltzmann factor $E_b/(k_B T) \propto 1/(T a^2)$. Therefore, an increase of the sample temperature would cause the same molecular fraction to form at smaller interaction strengths, corresponding to larger molecule binding energies $E_b$, and leading to $(T/T_F)_c \propto 1/(\kappa_F a)^2$. This is qualitatively different from the observed temperature dependence, displayed in Fig.~3d. While we cannot rule out a beneficial effect associated with the presence of molecules at the interface, which may partially lower the tendency of the atoms to recombine into dimers, we can rule out that dimer formation lies at the origin of the observed dynamical behaviour.

\section{Spin response on the lower (attractive) branch}
\begin{figure}[b!]
\centering
\includegraphics[width=9cm]{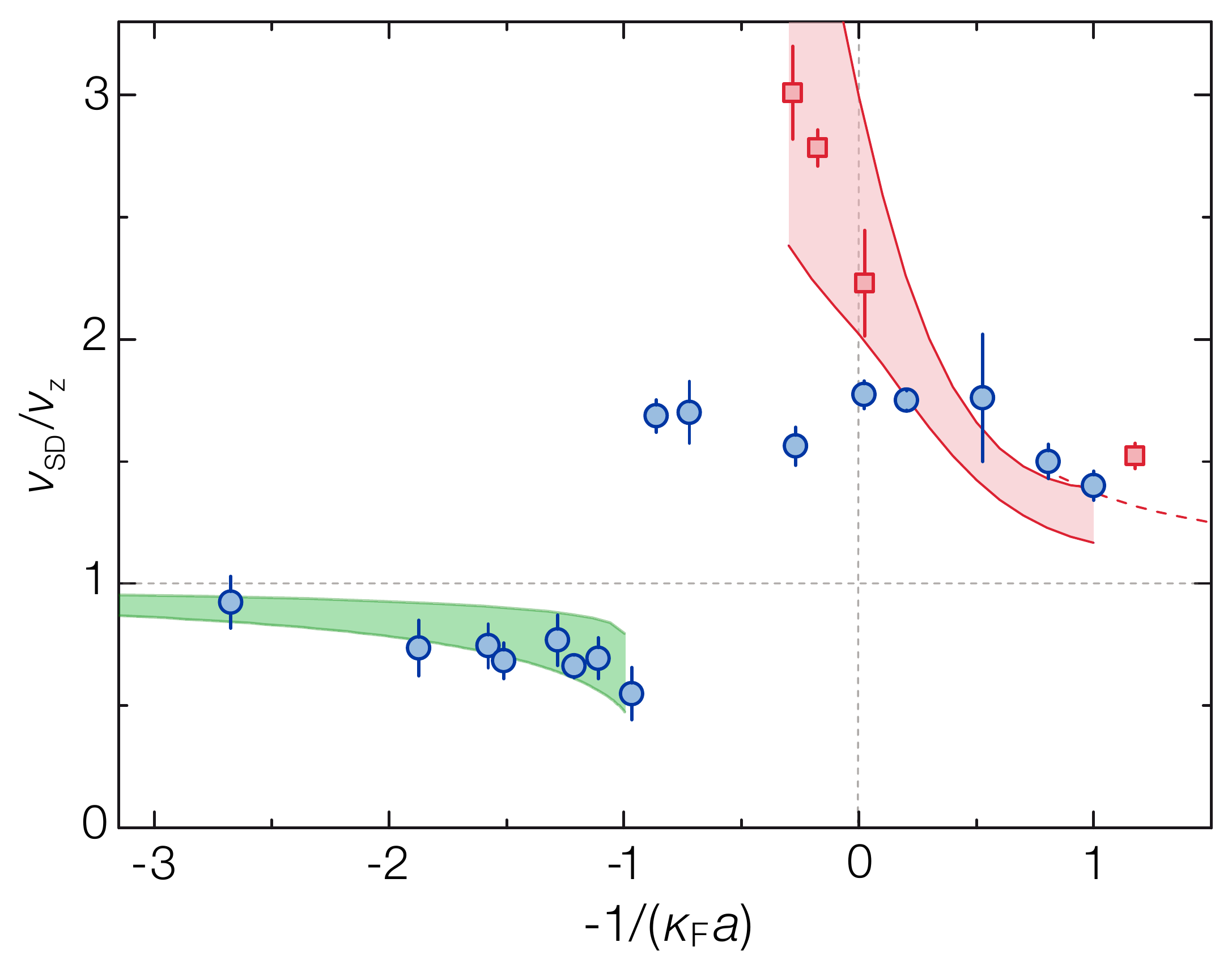}
\caption{Spin-dipole frequency of an attractive Fermi gas at the BEC-BCS crossover. Red squares: $\nu_{SD}/\nu_0$ obtained after intentionally initializing the system in the lower branch. Blue circles: spin-dipole frequency data shown in Fig.~2-3 of the main text, initializing the system to a repulsive Fermi gas. Green shaded region: theory prediction from Eq. (1) for a repulsive Fermi liquid, as shown in Fig. 2d of the main text. Red shaded region: theory prediction based on Eq. \eqref{chichi0} for an attractive Fermi gas at $T/T_F = 0.2$, obtained from the spin susceptibility and chemical potential given in Ref. \cite{Tajima2016}. The upper (lower) curve is obtained assuming full (20$\%$) spatial overlap between the two spin clouds. The dashed red line denotes the theoretical expectation based on first-order perturbation theory \cite{Recati2010}.}
\label{attr-spin-dipole}
\end{figure}

To validate further our interpretation of the spin-dipole oscillation data presented in Fig. 2 and 3 of the main text, we have performed additional spin response measurements, employing a system intentionally initialized in the lower energy branch. The results are displayed in Fig.~\ref{attr-spin-dipole}.
The importance of such additional measurements is two-fold: on the one hand, we obtain novel experimental hints on the trend of the lower-branch spin susceptibility via the same protocol employed for investigating the spin response in the upper branch; on the other hand, we can rule out that the trend of $\nu_{SD}/\nu_z$ measured at strong coupling and shown in Fig.~2 of the main text could be ascribed to spin dynamics in the lower branch.
From both experimental data and theoretical calculations \cite{Nascimbene2011,Sanner2011,Tajima2016}, the spin susceptibility of a Fermi gas on the lower branch of the Feshbach resonance is known to exhibit a monotonic decrease when passing from the BCS to the BEC side of the crossover. This is equally true for an attractive Fermi liquid and for a paired state, and for the latter both in the superfluid phase or in a pseudogap regime. In particular, while the spin susceptibility of a weakly attractive system in the BCS limit tends to the value $\chi_0$ of a non-interacting Fermi gas, it strongly decreases when crossing the resonance and moving towards the BEC limit. Similarly, for all $\kappa_F a$ values, $\chi$ is expected to monotonically decrease with decreasing temperature, especially if this drops below the critical temperature $T_c$ for superfluidity \cite{Tajima2016,Bulgac2013}.   
Correspondingly, Eq. (1) of the main text (Eq.~\eqref{chichi0} here) predicts $\nu_{SD}/\nu_z>1$ for an attractive Fermi gas at all temperatures and interaction strengths, and $\nu_{SD}$ should present a sharp increase when crossing the resonance from $a<0$ to the $a>0$ (or once the temperature drops below $T_c$).

This supplementary characterization of the system dynamics is carried out by means of a slightly different experimental protocol with respect to the one described in Section 1 and in the main text.
We initially prepare a deeply degenerate system at $T/T_F\simeq0.12$ in the usual domain-wall configuration, at a bias field $B_0= 910$\,G, corresponding to  $\kappa_F a \simeq -1$. Here, the measured spin-dipole frequency, though still exceeding $\nu_z$, lies well below the value of $\nu_{SD}=1.70(4)$ obtained at $\kappa_F a > 1$, and no arrest in the spin diffusion is ever observed. Consequently, we can assume that at this field, after the barrier removal, the gas immediately occupies the lower branch of the many-body system, as the upper one is neither well-defined nor energetically accessible.

Once the barrier is removed, we hold the field at $B_0$ for 20\,ms, allowing a small mixing of the two spin clouds, and we subsequently ramp it to a target value $B_{target}<B_0$ through a 25\,ms ramp. We have checked that such a ramp is sufficiently adiabatic, so that the subsequent spin dynamics depends only weakly on the ramp duration. 
Once $B_{target}$ is reached, we monitor the spin dynamics measuring the evolution of the relative distance between the centers-of-mass of the two spin clouds. Through the same analysis procedure already described in the main text and Section~1 of the Supplementary Information, we extract $\nu_{SD}/\nu_0$ as a function the target interaction strength, as shown in Fig.~\ref{attr-spin-dipole} (red squares).   
It is evident how the spin-dipole frequency measured after the adiabatic ramp from the BCS side, while being consistent with data shown in the main text for $-1/\kappa_F a>0.5$, presents a strong deviation further towards the BEC side. In this region, $\nu_{SD}$ sharply increases to values above $2\nu_0$ already at unitarity, exceeding $3\nu_0$ at $1/\kappa_F a=0.25$. Further decreasing $B_{target}$, hence $\kappa_F a$, no clear frequency can be extracted, owing to strong damping of the oscillations. Comparing the lower-branch data with the ones obtained by initializing the system in the repulsive state (blue circles in Fig.~\ref{attr-spin-dipole}) seems to indicate that the upper branch becomes completely unstable towards pairing beyond unitarity at $-1/\kappa_F a > 0.5$, such that the associated spin response is entirely compatible with the gas occupying the lower branch already at the beginning of the evolution.

The trend of our lower-branch data is well captured by Eq. \eqref{chichi0}, provided that we insert the spin susceptibility $\chi$ and chemical potential $\mu$ of a BEC-BCS crossover Fermi gas above the critical temperature for superfluidity. For our theory prediction (shown in Fig.~\ref{attr-spin-dipole} as the red shaded region), we have used recent data obtained within an extended T-matrix approximation and including pairing fluctuations \cite{Tajima2016}, which match both QMC results at unitarity \cite{Bulgac2013} and experimental measurements of the lower-branch spin susceptibility \cite{Sanner2011}. 

The validity of the theoretical treatment based on Eq.~\eqref{chichi0} is therefore successfully confirmed in the well-established case of a BEC-BCS crossover attractive Fermi gas and our Supplementary Data of $\nu_{SD}$  match previous experimental measurements of $\chi$ by Sanner~\textit{et al.}~\cite{Sanner2011}. 
Finally, the reasonable agreement we find between our theoretical prediction and experimental data suggests that our system is in a pseudo-gap regime of pre-formed pairs, rather than in an attractive Fermi liquid phase. For the latter case, in fact, both experimental investigation and QMC studies at zero temperature \cite{Nascimbene2011} reported a value for the spin susceptibility at unitarity that would lead to a spin-dipole frequency significantly lower than the one we measure. Such an intriguing aspect definitely will be the subject of future investigations.

\section{Theoretical model for the domain-wall melting}
\label{FM_melt}
In this Section, we describe the phenomenological model developed to explain the finite duration of the spin diffusion plateaus, as a function of interaction and temperature. Such a model is based on the precise knowledge of the spectral properties and lifetime of $\downarrow$ impurities embedded in a Fermi sea of $\uparrow$ particles \cite{Schmidt2011}.
As discussed in the main text, the spectral function of such system is characterized by an upper and a lower branch. Depending on the interaction strength, associated to the lower branch there exist in the extremely imbalanced case two kinds of Landau's quasi-particles, coined attractive polarons and dressed molecules \cite{Massignan2014}. For interaction parameters $1/(\kappa_F a)>0.9$ ($1/(\kappa_F a)<0.9$) dressed molecules (attractive polarons) represent the absolute ground state of the many-body system. 
The upper branch, in turn, features a third kind of quasi-particle, termed repulsive polaron, whose existence has been experimentally demonstrated both in a three-dimensional mass-imbalanced Fermi mixture of $^{40}$K and $^{6}$Li atoms \cite{Kohstall2012}, and also in a two-dimensional Fermi mixture of $^{40}$K atoms \cite{Koschorreck2012}.
While the lower branch is associated to a net attractive interaction between the impurity and the particles of the medium, the upper one requires repulsion between the two species in order to develop. For increasing repulsive interaction ($1/(\kappa_F a)\rightarrow 0^+$), the energy of the repulsive polaron progressively increases. However, parallel to this, the repulsive polaron acquires a progressively shorter lifetime, set by the tendency of the system to decay onto the lower-lying states of the attractive branch.  
The decay rate $\Gamma$ associated with such inelastic processes, together with the energies $E_+$ and $E_-$ of repulsive and attractive polarons, has been determined through non-perturbative theory approaches \cite{Schmidt2011, Massignan2014}.   

Although our system is a 50-50 balanced mixture of $\uparrow$ and $\downarrow$ fermions, the results obtained in the impurity limit are extremely relevant for understanding the existence of a plateau of zero diffusivity in the spin dynamics. Since our studies start by preparing a phase-separated state, rather than a mixed paramagnetic phase, the initial mixing processes at the interface between the two spin domains can be viewed as events in which fermions of one kind enter in the Fermi sea of the other component, and vice-versa. Those can be described in terms of dressed quasiparticle properties derived in the impurity limit. 

Our proposed explanation of the conductance plateau shown in Fig.~3 of the paper proceeds along the following line of thought. In the case of purely repulsive interactions, the ferromagnetic state, if energetically allowed, would be indefinitely stable, and in a system with separately fixed spin populations would correspond to a phase-separated state. In fact, the miscibility of the two components would be prevented by the existence of a domain wall: namely, a $\downarrow$ fermion at the interface would need to pay a finite amount of energy $\sigma>0$ in order to access the other spin domain, forming a repulsive polaron.
In our system, however, if a repulsive polaron is created, it can subsequently decay onto the lower branch with a rate $\Gamma$, releasing an energy equal to the mismatch between the two branches, $E_+-E_-$. Hence, this two-step process will cause a net increase of energy $\Delta E= E_+-E_- - \sigma$ at a rate $\Gamma$. 
We assume that at the beginning of the dynamics, the energy associated to the domain wall is given by $\sigma \times N_{int}$, $N_{int}$ being the total number of fermions within a slice around $z=0$ of total thickness equal to one interparticle spacing (see Section \ref{EFandkF}). The duration of the plateau $\tau_p$ is then set by the condition:
\begin{equation}
\sigma \, N_{int}= (E_+-E_- - \sigma)\, \tau_p\, \Gamma 	
\label{Eq1}
\end{equation}
  
We write $\sigma = E_+ - E_{+c}$, where $E_{+c}$ denotes the energy of one free fermion at the interface. In homogeneous systems and in the zero-momentum impurity limit \cite{Massignan2014}, $E_{+c}= E_F$. In our inhomogeneous sample at finite temperature, and in the vicinity of the interface between the two spin clouds, we introduce $E_{+c}$ as a phenomenological fitting parameter.
Consequently, from Eq.~\eqref{Eq1} we obtain the following expression for $\tau_p$:
\begin{equation}
\tau_p=  \frac{N_{int} (E_{+}-E_{+c})}{\Gamma (E_{+c}-E_{-})} \frac{1}{2 \pi \nu_F}	
\label{Eq2}
\end{equation}
where $h \nu_F=\epsilon_F$.
To compare the prediction of Eq.~\eqref{Eq2} to our data we employ the values of $E_{+}$, $E_{-}$ and $\Gamma$ given as a function of interactions by Schmidt \textit{et al.} \cite{Schmidt2011}, and the values of $\kappa_F$, $\epsilon_F$ and $N_{int}$ obtained after radial averaging over finite temperature density profiles as described in Section \ref{EFandkF}. 
 
For each temperature regime herein investigated, the only free parameter of the model, $E_{+c}$, is fixed by optimizing the agreement between experimental data and the prediction of Eq.~\eqref{Eq2}. $E_{+c}$ non-linearly increases upon increasing temperature: this is expected since $E_{+c}$ roughly corresponds to the energy of the repulsive polaron branch at the $\kappa_F a$ value at which a first non-zero $\tau_p$ is detected. It is worth stressing how such a simple theory model, which accounts for finite temperature effects only via $E_{+c}$ and a renormalization of $\kappa_F$, $\epsilon_F $ and $N_{int}$, is able to provide a quantitative description of the experimentally observed trend.

The model prediction stops at the unitary point and does not extend on the $a<0$ side of the Feshbach resonance, because the theory becomes unreliable in this region, the decay rate of the upper branch reaching the order of the Fermi energy, and hence making the repulsive polaron an ill-defined quasiparticle \cite{Schmidt2011}. As a matter of fact, we are at present unable to provide a description for such a feature (and for its disappearance) on the BCS side of the resonance, which we cannot exclude to arise from a combination of collisional and many-body effects. Certainly, the measured $\tau_p$ and $\nu_{SD}$ at low temperatures are consistent with the system temporarily accessing the upper branch even for $\kappa_F a<0$ \cite{Pekker2011}. This extremely exotic many-body state, which is a three-dimensional analogue of the super-Tonks regime in one dimension \cite{Haller2009}, is thus far poorly explored and definitively deserves deeper theoretical and experimental investigation.

\section{Spin drag coefficient and theory model for the diffusion of an impurity in a Fermi sea}\label{DiffPol}

The equation for the diffusion dynamics of the relative center of mass $d(t)=z_\uparrow - z_\downarrow$ can be easily obtained starting from the 
Boltzmann equation and using the method of the averages \cite{Vichi99}. One obtains the following coupled equations:
\begin{eqnarray}
 \partial_t(z_\uparrow - z_\downarrow)-(v_\uparrow - v_\downarrow)&=&0\nonumber\\
 \partial_t(v_\uparrow - v_\downarrow)+\omega_z^2(z_\uparrow - z_\downarrow)&=& (\partial_t(v_\uparrow - v_\downarrow))_{coll},
 \label{avmeth}
\end{eqnarray}
where $\omega_z$ is the trapping frequency in the direction of the motion, $v_\uparrow$ ($v_\downarrow$) is the average velocity of the $\uparrow$ 
($\downarrow$) spin component and $(\partial_t(v_\uparrow - v_\downarrow))_{coll}$ is the collisional term. The effect of the medium is included in 
this approach only in the collisional term. Assuming that the distribution function 
changes in time only through the change of velocity, the collisional term is simply proportional to the relative velocity and can be written as 
\begin{equation}
 (\partial_t(v_\uparrow - v_\downarrow))_{coll}=-\Gamma_S (v_\uparrow - v_\downarrow),
 \label{spindr}
\end{equation}
where $\Gamma_S$ is the so-called spin drag coefficient due to collisions \cite{Sommer2011,Enss2012}.
Therefore, substituting Eq.~\eqref{spindr} in Eq.~\eqref{avmeth} the equation of motion for $d$ is simply given by the one of a damped harmonic oscillator: 
\begin{equation}
 \ddot d+\Gamma_S\, \dot d+\omega_z^2 d=0
\label{diffeqn}
\end{equation}
We obtain the experimental spin drag coefficient by fitting the solution of Eq.~\eqref{diffeqn} to the data, with initial conditions $d(0)=d_0$ and ${\dot d}(0)=0$, or equivalently $\Delta M(0)=\Delta M_0$ and ${\dot \Delta M}(0)=0$ (see Fig.~\ref{diffusion}).

A main theoretical task is to determine the spin drag coefficient. For $\downarrow$ impurities moving through a fully polarized noninteracting $\uparrow$ Fermi sea, the drag rate can be
computed as
\begin{multline}
  \label{eq:dragrate}
  \hbar\Gamma_S = \frac{2\pi}{k_BTn_\downarrow}
  \int \frac{d^3p_1}{(2\pi\hbar)^3}\, \frac{d^3p_2}{(2\pi\hbar)^3}\,
  \frac{d^3p_3}{(2\pi\hbar)^3}\,
  \delta(\varepsilon_1+\varepsilon_2-
  \varepsilon_3-\varepsilon_4) \\
  \times \frac{(4\pi\hbar^2)^2}{m^2} \,
  \frac{d\sigma}{d\Omega} \,
  p_{1j}(v_{1j}-v_{3j})\, n_{1\downarrow} n_{2\uparrow}
  (1-n_{3\downarrow}) (1-n_{4\uparrow}).
\end{multline}
This collision integral describes the scattering of an impurity atom
with momentum $\vec p_1$ and a medium atom with momentum $\vec p_2$
into outgoing states $\vec p_3$ and
$\vec p_4=\vec p_1+\vec p_2-\vec p_3$, conserving the total momentum
and the total kinetic energy
$\varepsilon_1+\varepsilon_2$ of both particles.
Each particle has kinetic energy $\varepsilon_{\vec p}=p^2/2m$ and
mass $m$.  The drag rate is proportional to the change in impurity
velocity, $v_{1j}-v_{3j}$, where $j$ denotes the spatial component in
the direction of the initial impurity velocity.  The scattering
process occurs with probability $n_{1\downarrow} n_{2\uparrow}$ that
the initial states are occupied, and probability
$(1-n_{3\downarrow}) (1-n_{4\uparrow})$ that the final states are
unoccupied, where
$n_{\vec p\sigma} = [\exp(\beta(\varepsilon_{\vec
  p\sigma}-\mu_\sigma))+1]^{-1}$
is the Fermi-Dirac distribution at chemical potential $\mu_\sigma$.
The drag rate Eq.~\eqref{eq:dragrate} is derived from spin diffusion
theory for a polarized Fermi gas \cite{jeon1987, bruun2011,
  enss2013trans} in the limit of vanishing minority density
$n_\downarrow$ (the integral over the impurity distribution
$n_{1\downarrow}$ cancels the factor $n_\downarrow$ in the denominator
of Eq.~\eqref{eq:dragrate} to yield a finite drag rate).
Equivalently, Eq.~\eqref{eq:dragrate} is obtained from the impurity
drag rate \cite{bruun2008, Duine2010} in the limit of vanishing
impurity velocity.

In order to compute the cross section $d\sigma/d\Omega$ for a dilute
Fermi gas we use the T matrix in the ladder approximation \cite{bishop1976}.  
In an ultracold Fermi gas the bare interaction is
a $s$-wave contact interaction between unequal spins, thus also the T
matrix only has an $s$-wave component $\ell=0$.  One can then express
the differential cross section
\begin{align}
  \label{eq:crosssection}
  \frac{d\sigma}{d\Omega} = \lvert f_{\ell=0}(\vec q,\omega) \rvert^2
\end{align}
in terms of the $s$-wave scattering amplitude
$f_{\ell=0}(\vec q,\omega)$ of two incoming particles with total
momentum $\vec q=\vec p_1+\vec p_2$ and total kinetic energy
$\hbar\omega=\varepsilon_{\vec p_1\uparrow} + \varepsilon_{\vec
  p_2\downarrow}$.
The scattering amplitude, in turn, is given in terms of the T matrix
$\mathcal T_\ell$ as \cite{bishop1976}
\begin{align}
  \label{eq:scattampl}
  f_{\ell=0}(\vec q,\omega)
  = -\frac{mQ}{4\pi\hbar^2}\, \mathcal T_{\ell=0}(\vec q,\omega).
\end{align}
For two particles scattering in vacuum ($Q=1$), the two-body T matrix
reads
\begin{align}
  \mathcal T_{\ell=0}^{(0)}(\vec q,\omega)
  = \frac{4\pi\hbar^2a/m}{1+iak}
\end{align}
where $a$ denotes the $s$-wave scattering length, and
$\hbar\vec k=(\vec p_1-\vec p_2)/2$ is the relative momentum of
incoming particles.  This results in a vacuum scattering amplitude
$f_{\ell=0}^{(0)} = -a/(1+iak)$ and vacuum scattering cross section
$d\sigma^{(0)}/d\Omega=\lvert f_{\ell=0}^{(0)}\rvert^2 =
a^2/(1+a^2k^2)$.
At weak coupling $|k_{F\uparrow}a|\ll1$, the drag rate is proportional
to the scattering cross section, $\Gamma_S \propto a^2$.  Note that
with the vacuum cross section, the drag rate depends only on the
modulus $|a|$ of the scattering length and is symmetric in $a$ around
unitarity ($a^{-1}=0$) in the BCS-BEC crossover.  The experimental
data for the drag rate, however, exhibit a small asymmetry in $a$.  A
similar asymmetry in transport coefficients has been observed in the
shear viscosity \cite{Elliott2014} and transverse spin diffusion \cite{Trotzky2015}.

In order to explain the asymmetry in the drag rate it is necessary to
include medium scattering, where the Fermi sea is Pauli blocked for
intermediate states, and which entails a tendency toward molecule
formation on the BEC side \cite{bruun2005, Chiacchiera2009, enss2012crit,bluhm2014}.
Medium scattering is described by the many-body T matrix
$\mathcal T_\ell$,
\begin{align}
  \mathcal T_{\ell=0}^{-1}(\vec q,\omega)
  = \mathcal T_{\ell=0}^{(0)-1}(\vec q,\omega)
  + \int \frac{d^3p}{(2\pi\hbar)^3}\, \frac{n_{\vec p\downarrow} +
  n_{\vec q-\vec p\uparrow}} {\omega - \varepsilon_{\vec p\downarrow} -
  \varepsilon_{\vec q-\vec p\uparrow} +i0}\,.
\end{align}
The medium T matrix includes the effect of quantum degeneracy, which
leads to a large increase in the partial-wave scattering amplitudes
\eqref{eq:scattampl} and would by itself violate the unitarity bound
\begin{align}
  \label{eq:unibound}
  \lvert kf_\ell \rvert \leq 1,
\end{align}
which is the prerequisite for expressing the scattering amplitude
$kf_\ell = e^{i\delta_\ell} \sin \delta_\ell$ in terms of real phase
shifts $\delta_\ell(\vec q,\omega)$.  The definition of the scattering
amplitudes \eqref{eq:scattampl} in the presence of the medium
therefore includes a phenomenological $Q$ factor which accounts for
the fact that only unoccupied states are available for outgoing waves \cite{bishop1976}
(Pauli blocking),
\begin{align}
  \label{eq:Q}
  Q = \int_{-1}^1 \frac{d\cos\theta}{2} \left( 1 - n_{\vec q/2+\hbar\vec k\downarrow} -
  n_{\vec q/2-\hbar\vec k\uparrow} \right).
\end{align}
The integral averages over the angle $\theta$ between total momentum
$\vec q$ and relative momentum $\hbar\vec k$, while the modulus of $\vec k$
is fixed by the condition
$\hbar\omega = \varepsilon_{\vec q/2+\hbar\vec k\downarrow} +
\varepsilon_{\vec q/2-\hbar\vec k\uparrow} = \hbar^2k^2/m+q^2/4m$.
In vacuum $Q=1$, and also at high temperatures $T\gg T_{F\uparrow}$
one has $Q\approx 1$.  However, in a degenerate Fermi gas $Q<1$ lowers
the scattering amplitude sufficiently to always satisfy the unitarity
bound \eqref{eq:unibound} also in the case of medium scattering.

The drag rate Eq.~\eqref{eq:dragrate} is known analytically in the
whole BCS-BEC crossover in the high-temperature limit
$T\gg T_{F\uparrow}$ where the majority Fermi gas is non-degenerate \cite{enss2013trans},
\begin{align}
  \label{eq:draghightemp}
  \Bigl( \frac{\hbar \Gamma_S}{E_{F\uparrow}} \Bigr)_\text{hightemp}
  = \frac{16\sqrt2}{9\pi^{3/2}}
  \Bigl(\frac{T_{F\uparrow}}{T}\Bigr)^{1/2}
  \bigl[1-x-x^2e^x\,\text{Ei}(-x)\bigr]_{x=\hbar^2/(ma^2k_BT)}
\end{align}
where $\text{Ei}$ denotes the exponential integral.  In this regime,
the medium factor $Q=1$.  Conversely, medium scattering becomes
important in a degenerate Fermi gas ($T\lesssim T_{F\uparrow}$).  In
general, the collision term in Eq.~\eqref{eq:dragrate} can be reduced
to a three-dimensional integral which is readily evaluated
numerically.  The resulting drag rate exhibits a maximum slightly on
the BEC side of the resonance, in agreement with our experimental results.  As the
temperature is increased, the maximum shifts toward unitarity in
agreement with the high-temperature expression \eqref{eq:draghightemp}
which is symmetric in $a$. At very low temperatures the agreement between this theory prediction and the experimental data is poorer, since there the T-matrix approximation is known to be quantitatively incorrect. Moreover the gas may suffer some heating during the dynamics due to decay processes, making its temperature higher than the one measured at the start of the dynamics, which is used for the theory comparison. 

\section{Breathing mode characterization and collisional effects}
In this Section we elaborate on the role of collisional effects, which could in principle dominate the system collective dynamics leading to misinterpretation of the observed spin-dipole mode behaviour. In strongly interacting many-body systems, collective modes can be crucially affected by collisional hydrodynamics \cite{Chiacchiera2009}. Within a purely collisional framework, the amplitude of spin-dipole oscillations is governed by the differential equation \eqref{diffeqn}, which we use to analyze the long-time spin diffusion dynamics. Neglecting in-medium corrections, the collisional approach is based on the two-body scattering cross-section $\sigma(k)=4\pi a^2/(1+(k a)^2)$ for relative momentum $k$. This leads to predictions independent of the sign of $a$ and irrespective of whether the system is prepared on the upper or lower branch, in clear contrast with our observations summarized in Fig.~\ref{attr-spin-dipole}. This would not change if one takes in-medium corrections into account (see Section \ref{DiffPol}): at low temperatures these would lead only to a small asymmetry in the collisional integral \cite{Chiacchiera2009}, letting for instance many transport coefficients reach their maxima or minima slightly on the BEC side of resonance (see e.g. Fig~4 of the main text).

Let us now consider the upper branch data for $0<\kappa_Fa<1$: the collisional hydrodynamics prediction \eqref{diffeqn} for the spin-dipole frequency might appear qualitatively compatible with the observed softening, given that $\Gamma_S$ increases with interactions \cite{Goulko2011} (see Fig.~4 of the main text). As long as $\Gamma_S < 2\omega_z$, Eq.~\eqref{diffeqn} would indeed give a spin-dipole frequency $\omega_{SD} = (\omega_z^2-\Gamma_S^2/4)^{1/2}$. However, the damping rates $\Gamma_S$ experimentally extracted by fitting the long-time evolution with the full solution of Eq.~\eqref{diffeqn} greatly exceed $\omega_z$, yielding imaginary frequencies. 
On the other hand, we can consider Eq.~\eqref{diffeqn} as a phenomenological model for the spin-dipole oscillations. In this case, the damping rate is the value $1/\tau$ as extracted from the fitting procedure described in Section~\ref{Expcond}, and therefore the measured spin-dipole frequency should be $\omega_{SD}=(\omega_z^2-1/\tau^2)^{1/2}$. Our fitted parameters do not satisfy the previous relation and more generally a collisional 
picture. Indeed the fitted $\tau$ values (i) would give a very small correction to the bare frequency for any interaction strength and temperature and (ii) do not decrease 
by increasing the value of $\kappa_Fa$. 
Against a simple collisional interpretation speaks also the larger $\nu_{SD}$ found at higher temperature for fixed $\kappa_Fa<1$ (purple squares in Fig.~2d). In the collisional dynamics framework the oscillation frequency should instead decrease with temperature in the degenerate regime, where the damping rate increases because Pauli blocking becomes gradually less effective, as shown in Fig.~4.

\begin{figure}[t!]
\centering
\includegraphics[width=15cm]{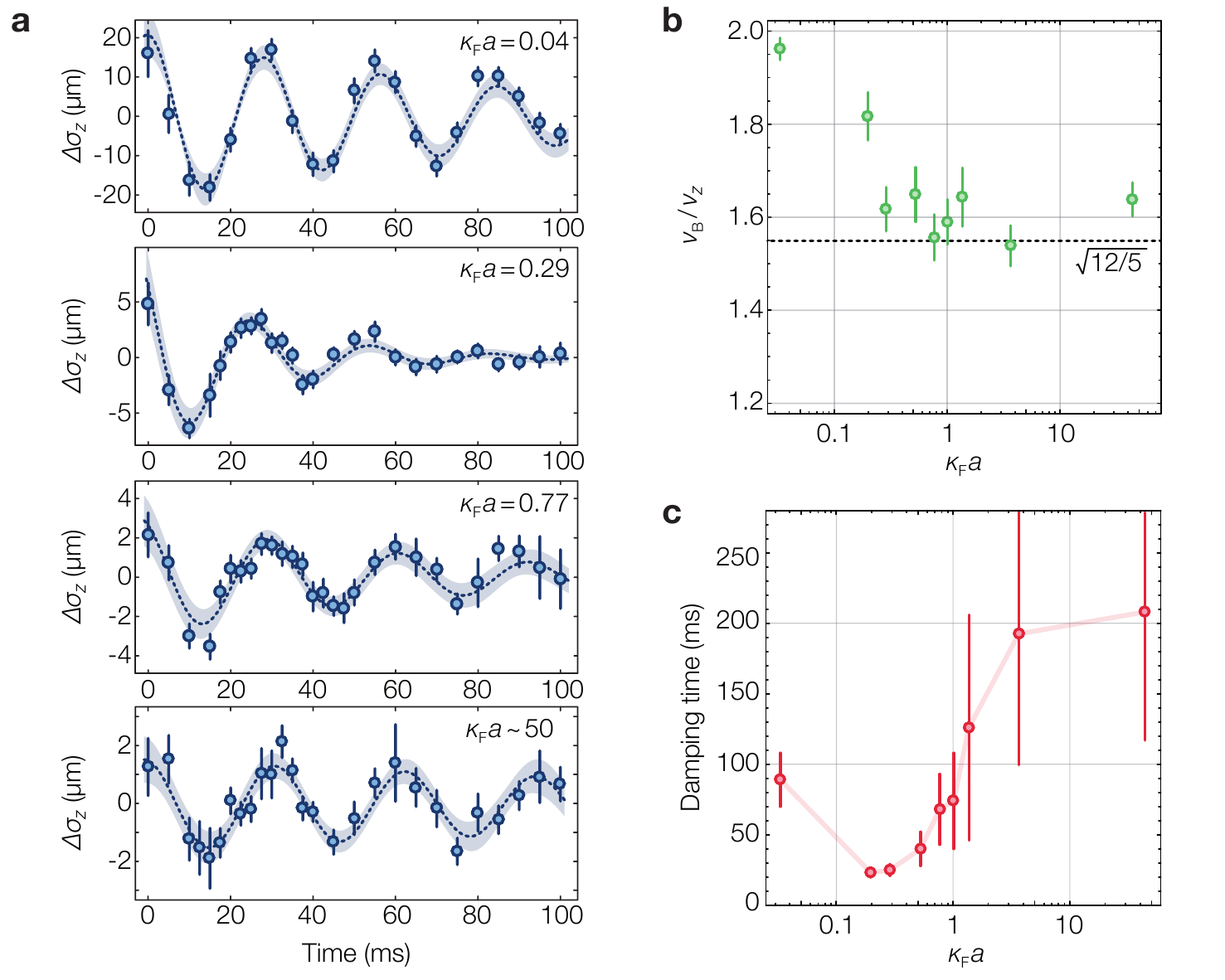}
\caption{Characterization of the axial breathing mode. 
\textbf{a}, Evolution of the axial width $\Delta \sigma_z (t)$ for different interaction strengths at $T/T_F=0.12(2)$ after a sudden barrier switch-off. \textbf{b}, Normalized breathing frequency as a function of $\kappa_F a$. The breathing mode frequency $\nu_B/\nu_z$ decreases from the collisionless value 2, reaching a value between $\sqrt{12/5}$ and $2$ already at $\kappa_F a \simeq 0.2$. \textbf{c}, Damping time of the breathing mode versus $\kappa_F a$. The transition from a collisionless to a collisional regime is additionally signalled by a minimum of the damping time at $\kappa_Fa \simeq 0.2$.}
\label{coll_to_hydro}
\end{figure}

\bigskip
\noindent\textbf{Breathing dynamics}
\newline
Besides exciting the spin-dipole mode, our experimental protocol also excites weakly the axial breathing mode, which we separately characterize by studying the evolution of the cloud axial width $\sigma_z$. Similarly to what we have done for the COM separation dynamics, we isolate the breathing oscillation by subtracting from $\sigma_z (t)$ an overall exponential drift and fitting the residual modulation $\Delta \sigma_z (t)$ with a damped sinusoidal function, as in Fig.~\ref{coll_to_hydro}a. The resulting frequency and damping time are displayed in Fig.~\ref{coll_to_hydro}b-c. For weak interactions, the axial breathing mode oscillates at twice the axial trap frequency $\nu_z$, as expected in the collisionless regime. Increasing the interaction strength, we observe a decrease of the breathing mode frequency $\nu_B$, which for $\kappa_F a \gtrsim 0.2$ reaches an approximately constant value slightly larger than the hydrodynamic expectation $\nu_B = \sqrt{12/5}\,\nu_z \simeq 1.55\,\nu_z$ \cite{Gensemer01} (see Fig.~\ref{coll_to_hydro}b). 

In contrast to Ref.~\cite{Sommer2011}, we find the breathing mode amplitude to be substantially smaller than the one of the out-of-phase COM oscillation at all interactions, owing to the small initial relative momentum imparted to the clouds by our experimental protocol. For this reason, the spin-dipole mode is largely decoupled from the breathing one as long as the two clouds do not start bouncing off each other, as if they became impenetrable. 
In this latter regime, observed at strong repulsion, the two modes are coupled and possibly lock to a common frequency within the experimental uncertainty. Such a frequency $\nu_{SD} \simeq 1.70(4)\,\nu_z$ is compatible to $\nu_R = 1.78(5)\,\nu_z$ that we measured for a spin-polarized cloud oscillating in ``half'' the trap when the optical barrier is left on; it is also similar to the experimental value $1.63\,\nu_z$~\cite{Sommer2011}, and especially to the theory prediction $1.78\,\nu_z$ \cite{Taylor2011}. The value is also close to the frequency $\sqrt{12/5}\,\nu_z$ of the breathing mode in the hydrodynamic regime. In Ref.~\citenum{Goulko2011}, where a purely collisional approach is developed for two colliding clouds, it is found that the breathing mode remains essentially collisionless as long as the two clouds can diffuse into each other rapidly.  
Conversely, when the diffusion is slow, the breathing and the spin-dipole modes become coupled and the two clouds are predicted to enter into a bouncing regime. In this regime, both modes are locked to the same frequency which coincides at low temperature with the hydrodynamic prediction $\sqrt{12/5}\,\nu_z$ \cite{Goulko2011}. In our experiment, the bouncing is observed only deep in the collisional regime, and therefore our measurements of the dipole and the breathing modes do not fit even qualitatively to a purely collisional approach. We further note that in the bouncing regime the two modes start being locked irrespective of the mechanism that makes the two clouds almost immiscible (see also Ref.~\citenum{Taylor2011}).

Furthermore, ascribing the bouncing dynamics to a collisional origin is completely inconsistent with the distinct behaviour of the spin-dipole frequency measured on the lower branch, where $\nu_{SD}$ is larger (smaller) than $1.55 \nu_z$ at $\kappa_F a \simeq 2$ ($\kappa_F a \simeq -1$). The characterization of the breathing mode, together with the measured lower-branch spin-dipole frequency (see Fig.~\ref{attr-spin-dipole}), lead us therefore to attribute the sudden jump of $\nu_{SD}$ in Fig.~2d to a genuine system immiscibility, ruling out a dynamical origin.


\end{document}